\documentclass[pra, twocolumn, amsmath, amssymb, notitlepage, longbibliography]{revtex4-1}

\usepackage[english]{babel}
\usepackage{graphicx}
\usepackage{hyperref}

\newcommand{\bra}[1]{\ensuremath{\left\langle #1\r|}}
\newcommand{\ket}[1]{\ensuremath{\left|#1\r\rangle}}

\newcommand{\mean}[1]{\ensuremath{\left\langle #1\r\rangle}}

\newcommand{\cc}{^{\ast}}                						
\newcommand{\hc}{^{\dagger}}             					
\newcommand{\op}[1]{\hat{#1}}							
\newcommand{\ee}{\mathrm{e}}             					
\newcommand{\ii}{\mathrm{i}}			             					
\renewcommand{\H}[0]{H}  							
\renewcommand{\L}[0]{\mathcal{L}}  						

\newcommand{\comm}[2]{\left[ #1, #2 \right]} 				
\newcommand{\nn}{\nonumber}							
\newcommand{\abs}[1]{\ensuremath{ \left| #1 \right| }}		
\newcommand{\abss}[1]{\ensuremath{ \left| #1 \right|^{2} }}	
\newcommand{\diss}[1]{\mathcal{D}[ #1 ]}					
\renewcommand{\l}[0]{\left}
\renewcommand{\r}[0]{\right}

\newcommand{\eq}[1]{\eqref{#1}}
\newcommand{\eqn}[1]{Eq.~\eq{#1}}

\begin{document}

\title{Dissipative Rabi model in the dispersive regime}

\author{Clemens M\"uller}
\email{clm@zurich.ibm.com}
\affiliation{IBM Quantum, IBM Research - Z\"urich, 8803 R\"uschlikon, Switzerland}
\affiliation{ARC Centre of Excellence for Engineered Quantum Systems, School of Mathematics and Physics, University of Queensland, Saint Lucia, Queensland 4072, Australia}


\date{\today}

\begin{abstract}
	The dispersive regime of circuit QED is the main workhorse for today’s quantum computing prototypes based on superconducting qubits. 
	Analytic descriptions of this model typically rely on the rotating wave approximation of the interaction between the qubits and resonators, 
	using the Jaynes-Cummings model as starting point for the dispersive transformation. 
	Here we present analytic results on the dispersive regime of the dissipative Rabi model, without taking the rotating wave approximation of the underlying Hamiltonian.
	Using a recently developed hybrid perturbation theory based on the expansion of the time evolution on the Keldysh contour~[Phys. Rev. A 95, 013847 (2017)], 
	we derive simple analytic expressions for all experimentally relevant dynamical parameters like dispersive shift and resonator induced Purcell decay rate, 
	focusing our analysis on generic multi-level qubits like the transmon.
	The analytical equations are easily tractable and reduce to the known Jaynes-Cummings results in the relevant limit. 
	They however show qualitative differences at intermediate and large detuning, allowing for more accurate modelling of the interaction between superconducting qubits and resonators. 
	In the limit of strong resonator driving, our results additionally predict new types of drive induced qubit dissipation and dephasing, not present in previous theories.
\end{abstract}


\maketitle

\section{Introduction}
	
	Quantum optics and the related fields of quantum computation are at their heart concerned with the interactions between atoms and light - 
	light as in coherent modes of electromagnetic radiation and atoms as in well controlled, engineered quantum few-level systems~\cite{QuantumOpticsBook}. 
	Typically this interaction is described in the so-called Jaynes-Cummings model~\cite{JaynesCummings, Blais:PRA:2004},
	which is based on an approximation of the more fundamental Rabi model~\cite{RabiModel},
	describing the interaction between the light and dipole allowed transitions in the atom.
	The differences between Rabi and Jaynes-Cummings model are thought to be small in the usual experimental situations, 
	and to see noticeable effects one usually needs to consider the ultra-strong coupling regime~\cite{FornDiaz:NP:2016, Kockum:NR:2019}, 
	where the coupling strength between atoms and light becomes comparable to their characteristic energy scales.
	In this paper we show that already for much smaller coupling strength, qualitative and quantitative differences between the two models can be observed.
	
	On the way towards further improving the degree of control over quantum hardware, we continuously need to improve our understanding of the quantum systems and their coupled dynamics. 
	Of special importance for quantum computation is the so-called dispersive regime of the atom-light interaction~\cite{Blais:PRA:2004}, 
	where the atomic transitions are detuned from the mode energies of the light field by more than the strength of their coupling. 
	It is in this regime that the successful early quantum computing prototypes based on 
	superconducting artificial atoms and microwave resonators are operated~\cite{Chow:NC:2014, Barends:N:2016, Roushan:S:2017, Kandala:N:2019}.
	It allows the resonator mode on one hand to isolate the sensitive qubits from most of the electromagnetic environment 
	while at the same time acting as an access port for qubit manipulation and state readout~\cite{QuantumEngineer}. 
	Similar advantages are also envisioned for coupling quantum dot spin qubits to microwave resonators~\cite{Petersson:N:2012, vanWoerkom:PRX:2018, Landig:NC:2019}.
	When modelling the system dynamics in the dispersive regime, one typically employs analytic approximations for the coupling-induced changes to Hamiltonian parameters 
	as well as for dissipative processes arising due to the atom-light hybridisation~\cite{Blais:PRA:2004, Boissonneault:PRA:2008, Boissonneault:PRA:2009}. 
	These, however, are usually based on the approximate Jaynes-Cummings interaction instead of the more fundamental Rabi model.
	
	Here, we show analytic expressions for Hamiltonian corrections and dominant dissipative dynamical contributions arising in the dispersive regime of the Rabi model directly, 
	without making the rotating wave-approximation underlying the Jaynes-Cummings interaction.
	Our treatment is based on a Keldysh diagrammatic perturbation approach~\cite{Mueller:PRA:2016}, 
	and delivers well-behaved and simple analytic expressions for all relevant parameters without requiring additional approximation.
	These results are relevant for more accurate analytical modelling of any quantum hardware in the dispersive regime, like superconducting and spin qubits. 
	The dissipative corrections to the dynamics we derive here are additionally important for a better understanding of qubit readout, when the resonators may be strongly driven.
	In this case, they lead to photon number dependent dissipative corrections to the qubit dynamics.

\section{Multi-level dissipative Rabi model \label{sec:Model}}

	We focus our treatment on a system of a multi-level atom coupled to a single, quantized resonator mode.
	This model accurately describes the typical situation for superconducting transmon qubits coupled to resonators in the circuit QED architecture. 
	Results for the special case of a two-level atom, more relevant to e.g. quantum-dot spin or charge qubits are detailed in Appendix~\ref{app:TwoLevel}.
	
	We write the total system Hamiltonian as $\H = \H_{0} + \H_{\text{int}} + \H_{\text{env}}$, with the unperturbed Hamiltonian
	\begin{align}
		\H_{0} = \omega_{r} a\hc a + \sum_{k=0}^{N-1} \omega_{k} \sigma_{k,k} \,.
		\label{eq:H0}
	\end{align}
	Here, we use the notation $\sigma_{k,l} = \ket k \bra l$ with $\ket k$ an eigenstate of the $N$-level system with eigenenergy $\omega_{k}$.
	$a$ is the bosonic annihilation operator of the resonator mode at frequency $\omega_{r}$.
	The interaction between multi-level qubit and resonator is written in the Rabi model as
	\begin{align}
		\H_{\text{int, Rabi}} = \sum_{k} g_{k} \l( \sigma_{k,k+1} + \sigma_{k+1,k} \r) \l( a\hc + a \r) \,,
	\end{align}
	assuming purely transversal coupling between the qubit and resonator, as is natural for transmon qubits. 
	Alternatively we could make a rotating wave approximation in the coupling, leading to the Jaynes-Cummings model interaction
	\begin{align}
		\H_{\text{int, JC}} = \sum_{k} g_{k} \l( \sigma_{k,k+1} a\hc + \sigma_{k+1,k} a \r) \,,
	\end{align}
	which discards fast rotating terms.
	
	The system is additionally coupled to three independent baths via $H_{\text{env}} = H_{\text{sys-env}} + H_{\text{env}, 0}$.
	The system-bath interactions are described by
	\begin{align}
		\H_{\text{sys-env}} =& \sum_{k} \beta_{k} \l( \sigma_{k,k+1} + \sigma_{k+1,k} \r)  \op X + \sum_{k} \delta\omega_{k} \sigma_{k,k} \op Z \nn\\
			&+ \l( a\hc + a \r) \op R \,.
	\end{align}
	Here the first term induces transitions in the multi-level system, at lowest order leading to energy dissipation and excitation.
	For weakly anharmonic qubits like the transmon, one usually finds the coupling strengths $\beta_{k} \sim \sqrt{k+1}$, but we do not restrict our analysis to this special case.
	The second term describes fluctuations of the qubits energy levels, leading primarily to qubit dephasing.
	The $\delta\omega_{k}$ characterise the sensitivity of the qubit level energy $\omega_{k}$ to small fluctuations in the external parameter $\op Z$.
	The final term then is to lowest order responsible for decay and excitation of resonator photons.
	
	We are assuming the bath operators $\op X, \op Z$, and $\op R$ to be hermitian operators of bosonic environments which are in thermal equilibrium at temperature $T$.
	The Hamiltonian $H_{\text{env},0}$ describes the internal dynamics of the baths, which we do not state here explicitly.
	More details on how we treat the bath operators can be found in Appendix~\ref{app:Bath}.
	
	{At this point it is instructive to emphasize the differences between the present model and our earlier work, Ref.~\cite{Mueller:PRA:2016}, which focused on a two-level double-quantum dot coupled to a microwave resonator and a phononic environment.
	In this work, the focus is instead on the multi-level nature of the qubit and the presence of multiple independent environments coupled to both the qubit and the resonator, leading to correlated dissipative processes like Purcell decay.}

\section{Perturbative treatment in the dispersive limit \label{sec:Results}}

	We use the technique developed in Ref.~\onlinecite{Mueller:PRA:2016} to derive effective dissipative rates and coherent Hamiltonian corrections 
	in the dispersive limit where $g\ll \abs{\Delta_{0}} = \abs{\omega_{1,0} - \omega_{r}}$. 
	This technique is based on a perturbative expansion of the density matrix time-evolution on the Keldysh contour. 
	The individual terms in this expansion can then be expressed as Keldysh diagrams, in the same spirit as Feynman diagrams for quantum state evolution, 
	and the resulting master equation can be written in Lindblad form.	
	Here we do a simultaneous perturbation theory in both the system interaction $H_{\text{int}}$ as well as the bath interaction $H_{\text{sys-env}}$, 
	assuming both interactions to be weak compared to the internal system dynamics. 
	The Keldysh expansion is well behaved also when considering the full Rabi interaction, and does not rely on the rotating wave approximation of the Jaynes-Cummings model.
	Then, at second order in the perturbation theory, we obtain the usual incoherent contributions to the dynamics, i.e. qubit dissipation and dephasing as well as resonator dissipation, 
	due to their coupling to their individual environments.
	Additionally, from the perturbative expansion including the atom-resonator interaction term $\H_{\text{int}}$, 
	we find corrections to the unperturbed Hamiltonian $H_{0}$ at second order, which we identify with the dispersive shifts~\cite{Blais:PRA:2004, Stace:PRL:2013}. 
	One of the major advantages of the Keldysh approach is that it provides a clear recipe for going to higher orders in perturbation theory.
	At fourth order this theory predicts, amongst others, 
	correlated dissipative processes that arise due to the hybridisation of the qubit and resonator states and their individual coupling to environments, which are at the focus of this paper. 
	Details on the technique and derivation of the master equation can be found in our earlier work~\cite{Mueller:PRA:2016}.
	
	We write the master equation resulting from the Keldysh diagrammatic expansion up to fourth order as
	\begin{align}
		\dot \rho = -\ii \comm{\H_{0} + H_{2}}{\rho(t)} + \mathcal L_{2} \rho(t) + \mathcal L_{4} \rho(t) \,.
		\label{eq:ME}
	\end{align}
	Here, $\H_{2}$ is a dispersive Hamiltonian correction 
	and $\L_{2}\rho$ summarises the incoherent dynamics at the same order.
	Finally $\L_{4}\rho$ contains all incoherent contributions at fourth order perturbation theory, which are the main focus of this work. 
	In the following we will provide the analytic expressions for all terms in Eq.~\eqref{eq:ME} and will 
	contrast the results when performing the perturbation theory either with the full Rabi model or the approximate Jaynes-Cummings interaction.

	\subsection{Second order dissipative terms}
	
	The Keldysh expansion at second order in the system-bath interaction $\H_{\text{sys-env}}$ reproduces the well-known incoherent terms in the master equation, which we reproduce here for clarity. 
	For a multi-level qubit and resonator, these are
	\begin{align}
		\L_{2} \rho(t) = \sum_{k}& \gamma_{\downarrow_{k}} \diss{\sigma_{k,k+1}}\rho(t) + \gamma_{\uparrow_{k}} \diss{\sigma_{k+1,k}}\rho(t) \nn\\
			&+ \gamma_{\varphi_{k}}\diss{\sigma_{k,k}}\rho(t) \nn\\
			&+ \kappa_{-}\diss{a}\rho(t) + \kappa_{+}\diss{a\hc}\rho(t) \label{eq:Diss2}
	\end{align}
	with the rates
	\begin{align}
		\gamma_{\downarrow_{k}} &= \beta_{k}^{2} C_{\op X}(\omega_{k+1,k})  \quad\,,\quad \gamma_{\uparrow_{k}} = \beta_{k}^{2} C_{\op X}(-\omega_{k+1,k}) \,,\nn\\
		\gamma_{\varphi_{k}} &= \delta\omega_{k}^{2}  C_{\op Z}(0) \,,\nn\\
		\kappa_{-} &= C_{\op R}(\omega_{r}) \quad\,,\quad \kappa_{+} = C_{\op R}(-\omega_{r}) \,.
	\end{align}
	Here the first line in Eq.~\eqref{eq:Diss2} describes qubit decay and excitation, the second line its pure dephasing of the qubit 
	and the final two terms incoherent photon loss and excitation from the resonator.
	We used the usual dissipator notation, $\diss{\op o}\rho = \op o \rho \op o\hc -\frac12\l( \op o\hc \op o \rho + \rho \op o \hc \op o \r)$.
	The spectral functions of the environmental operators we defined through $C_{\op o} (\omega) = \frac12 \int dt\: \ee^{\ii \omega (t-t')} \mean{\op o(t) \op o(t')}$.
	They describe the ability of the environment to exchange photons at energy $\omega$ with the system. More details in Appendix~\ref{app:Bath}.
	
	\subsection{Second order Hamiltonian corrections - dispersive shifts}
	
	\begin{figure}[htbp]
		\begin{center}
			\includegraphics[width=.95\columnwidth]{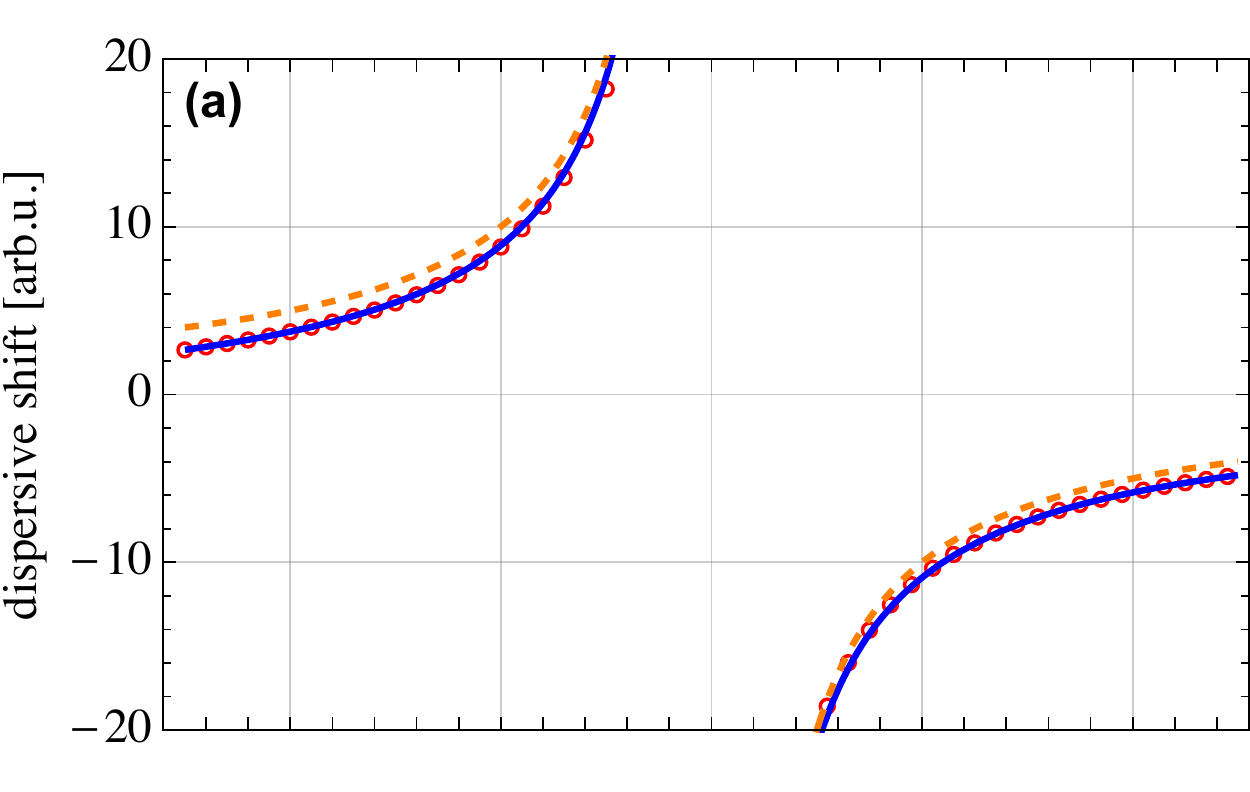}
			\includegraphics[width=.95\columnwidth]{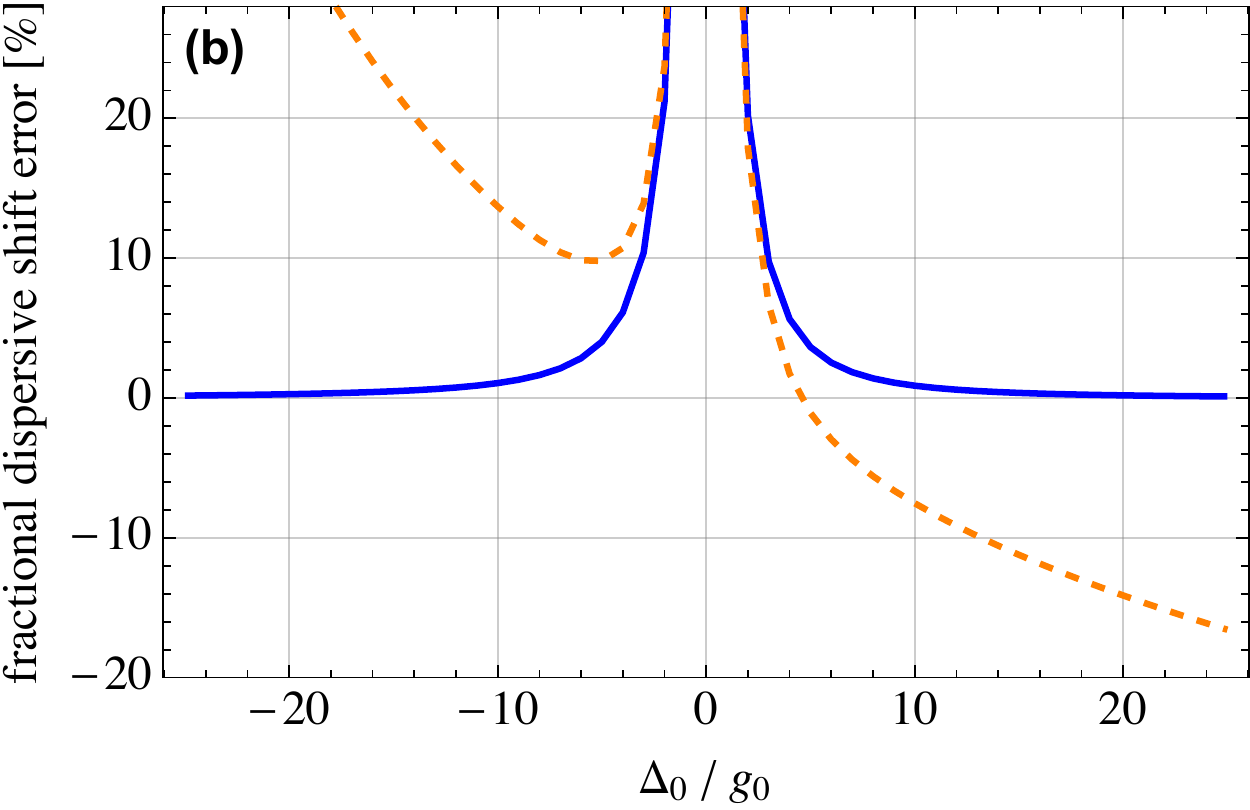}
			\caption{\textbf{(a)} Dispersive Hamiltonian corrections to the one resonator photon energy, $-\tilde\chi_{0}$ for the Rabi model (solid blue) 
				and $-\chi_{0}$ in the Jaynes-Cummings approximation (dashed orange), as function of detuning between resonator and lowest qubit transition,
				$\Delta_{0} = \omega_{1,0} - \omega_{r}$.
				Red circles are results from numerical exact diagonalisation of the Hamiltonian. 
				\textbf{(b)} Error of the dispersive shift as a fraction of the exact diagonalisation result.
				Parameters in this and all following plots: $\omega_{r}/2\pi = 5$~GHz, $g_{0}/2\pi = 100$~MHz.
				}
			\label{fig:DispersiveShift}
		\end{center}
	\end{figure}
	Additionally, our technique produces corrections to the Hamiltonian, due to the perturbative expansion in the system coupling operator $\H_{\text{int}}$.
	When taking the Rabi-type coupling $\H_{\text{int,Rabi}}$ as the perturbative interaction, we find these Hamiltonian correction at second order as
	\begin{align}
		\H_{2, \text{Rabi}} 
			=& \sum_{k} \l( \tilde\chi_{k-1} - \tilde\chi_{k} \r) a\hc a \sigma_{k,k} + \l( \chi_{k-1}  - \xi_{k} \r) \sigma_{k,k}\,,
	\end{align}
	with the generalized dispersive shift 
	\begin{align}
		\tilde\chi_{k} = \frac{2 g_{k}^{2}\:\omega_{k+1,k}}{(\omega_{k+1,k}^{2}-\omega_{r}^{2})}	
		\label{eq:Rabi}
	\end{align}
	and where we defined the qubit level splittings $\omega_{k+1, k} = \omega_{k+1} - \omega_{k}$, and $g_{k}=0\,, \forall k<0$.
	Here we also introduced the usual multi-level Jaynes-Cummings dispersive shifts $\chi_{k}$~\cite{Blais:PRA:2004} as well as the Bloch-Siegert shifts $\xi_{k}$~\cite{FornDiaz:PRL:2010} as
	\begin{align}
		\chi_{k} &= \frac{g_{k}^{2}}{(\omega_{k+1,k} - \omega_{r})} \,, \nn\\
		\xi_{k} &= \frac{g_{k}^{2}}{(\omega_{k+1,k} + \omega_{r})}\,.
	\end{align}
	Note that $\tilde\chi_{k} = \chi_{k} + \xi_{k}$.
	For the special case of a two-level atom, this result has previously been derived using different theoretical techniques~\cite{Zueco:PRA:2009, Zhu:NJP:2013}, see also Appendix~\ref{app:TwoLevel}.
	If we perform the perturbation expansion with the qubit-resonator interaction term in the Jaynes-Cummings approximation, $\H_{\text{int,JC}}$, we instead find
	\begin{align}
		\H_{2,JC} = \sum_{k} \l( \chi_{k-1} - \chi_{k} \r) a\hc a \sigma_{k,k} + \chi_{k-1} \sigma_{k,k} \,,
	\end{align}
	as is known in the literature~\cite{Boissonneault:PRA:2008, Slichter:PRL:2012}.
	Fig.~\ref{fig:DispersiveShift} shows a comparison of the Rabi and Jaynes-Cummings dispersive corrections to the one photon energy{, i.e. the energy required to add one photon to the resonator while the qubit rests in its groundstate. 
	These are given for the Rabi and Jaynes-Cummings model by $\tilde\chi_{0}$ and $\chi_{0}$, respectively.
	We show them }together with results from exact numerical diagonalisation of the Hamiltonian{ for the same parameters}, truncated to {include the ten lowest states for both multi-level qubit and resonator, for details see Appendix~\ref{app:MoreResults}}.
	The major qualitative difference is the asymmetry in the magnitude of the shift for positive and negative detuning in the result for the Rabi interaction~\cite{Zueco:PRA:2009, Zhu:NJP:2013}, 
	compared to the symmetric response of the Jaynes-Cummings result. 
	As a result, apart from a small region around small positive detuning, the error in the Rabi dispersive corrections is significantly smaller than for the Jaynes-Cummings result, 
	and Eq.~\eqref{eq:Rabi} converges to the exact results for large detuning. 
	For further comparison, including studying the behavior when fitting to experimental data and the implications for predictions of system dynamics, 
	see Appendix~\ref{app:MoreResults}.

	\subsection{Fourth order dissipative terms}
	
	At fourth order, a large number of dissipators contribute to the master equation. 
	Here we focus only on a subset of all incoherent contributions, and write the fourth-order dissipative contributions to the master equation as
	\begin{align}
		\mathcal L_{4}\rho = \mathcal L_\text{Purcell} \rho + \mathcal L_{\updownarrow,\pm} \rho + \mathcal L_{\varphi,\pm} \rho \,.
	\end{align}
	These three terms correspond to previously known corrections to the dissipative dynamics of a qubit and resonator in the dispersive regime~\cite{Mueller:PRA:2016, Blais:PRA:2004, Boissonneault:PRA:2008}.
	They arise from hybridisation of qubit and resonator states and the coupling of the resulting hybrid states to the qubit and resonator environments.  
	In the Keldysh perturbation theory they appear at fourth order, where two interaction vertices originate from the qubit-resonator interaction $\H_{\text{int}}$ 
	and the other two from the interactions with the baths $\H_{\text{sys-env}}$.
%
	In a previous work~\cite{Mueller:PRA:2016}, this subset of incoherent contributions has been shown to be dominating the steady-state properties in the dispersive parameter regime. 
	

	\subsubsection{Purcell process}
	
\begin{figure}[htbp]
		\begin{center}
			\includegraphics[width=.95\columnwidth]{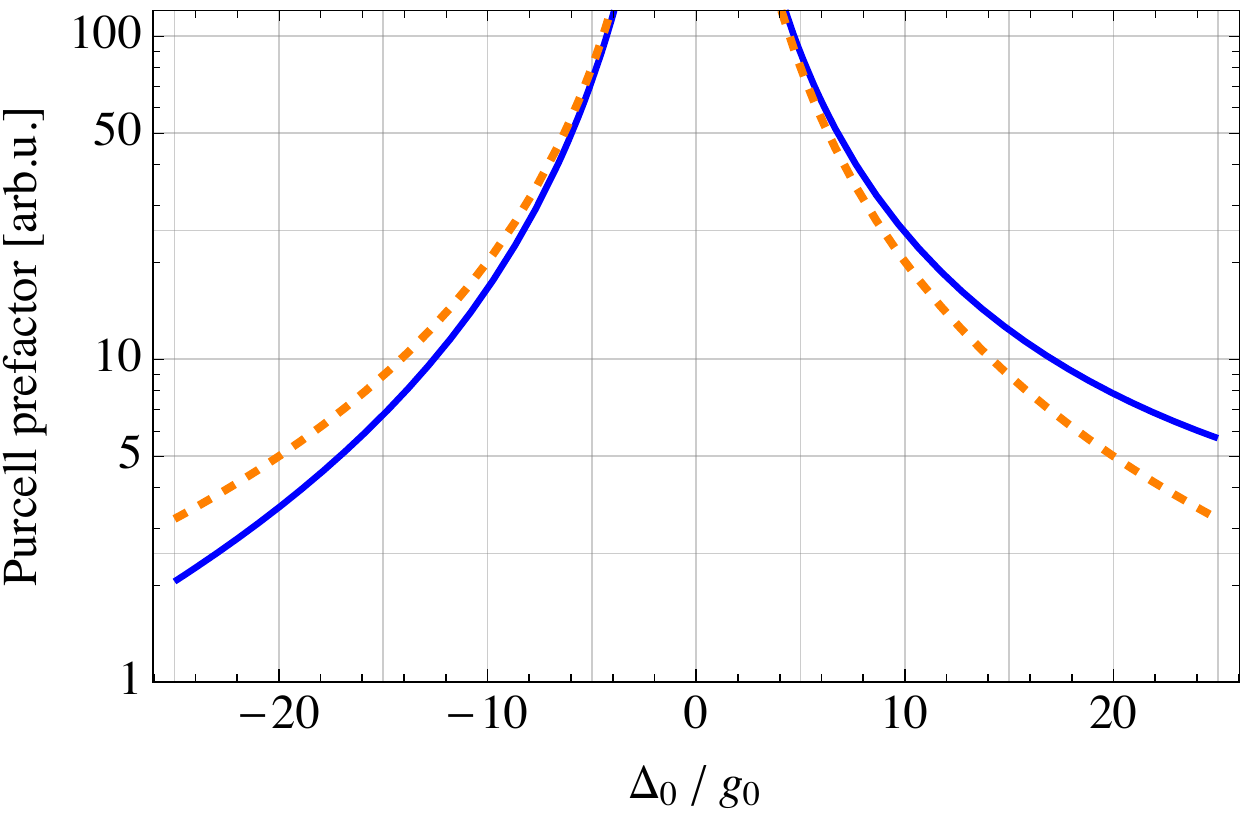}
			\caption{Comparison of the Purcell rate prefactor $p_{0}$, Eqs.~\eqref{eq:PurcellPreRabi},~\eqref{eq:PurcellPreJC}, 
				as function of detuning $\Delta_{0}$,  
				when taking the Rabi (solid blue) or Jaynes-Cummings interaction (dashed orange). 
				Note the asymmetry with respect to zero detuning for the Rabi result. 
				}
			\label{fig:Purcell}
		\end{center}
	\end{figure}

	The Purcell process describes decay and excitation of the qubit due to its effective coupling to the resonator environment. 
	We find its contribution to the master equation as
	\begin{align}
		\mathcal L_\text{Purcell} \rho 
			= \sum_k \gamma_{\text P,\downarrow}^{(k)} \mathcal D\left[ \sigma_{k,k+1} \right] \rho + \gamma_{\text P,\uparrow}^{(k)} \mathcal D\left[ \sigma_{k+1,k} \right] \rho \,,
	\end{align}
	with the Purcell decay and excitation rates
	\begin{align}
		\gamma_{\text P,\downarrow}^{(k)} &= p_{k} \: C_{\op R} \left(\omega_{k+1,k}\right) \,, \quad 
		\gamma_{\text P,\uparrow}^{(k)} = p_{k} \: C_{\op R} \left(- \omega_{k+1,k}\right) \,,
	\end{align}
	where we defined dimensionless prefactors $p_{k}$. For the Rabi model, these are
	\begin{align}
		p_{k,\text{Rabi}} = \frac{8 g_{k}^2 \:\omega_r^2}{(\omega_r^2 - \omega_{k+1,k}^2)^2} \,,
		\label{eq:PurcellPreRabi}
	\end{align}
	while for the Jaynes-Cummings interaction we find the canonical result
	\begin{align}
		p_{k,\text{JC}} = \frac{2g_{k}^2 }{(\omega_r - \omega_{k+1,k})^2} \,.
		\label{eq:PurcellPreJC}
	\end{align}
	Fig.~\ref{fig:Purcell} shows a comparison of the prefactors for the Rabi and Jaynes-Cummings model. 
	Similar to the dispersive shift, the inclusion of the fast rotating terms in the Rabi interaction leads to a notable asymmetry in the rates with respect to zero detuning. 
	For large negative detuning, $\omega_{1,0} < \omega_{r}$, Purcell is suppressed strongly compared to the Jaynes-Cummings result 
	while for large positive detuning, $\omega_{1,0} > \omega_{r}$ it is similarly enhanced.
	Note that the Purcell rates are proportional to the resonator environmental correlation function, $C_{{\op R}}(\omega)$, probed at the qubit transition frequencies.

	\subsubsection{Dressed dephasing}
	
\begin{figure}[htbp]
		\begin{center}
			\includegraphics[width=.95\columnwidth]{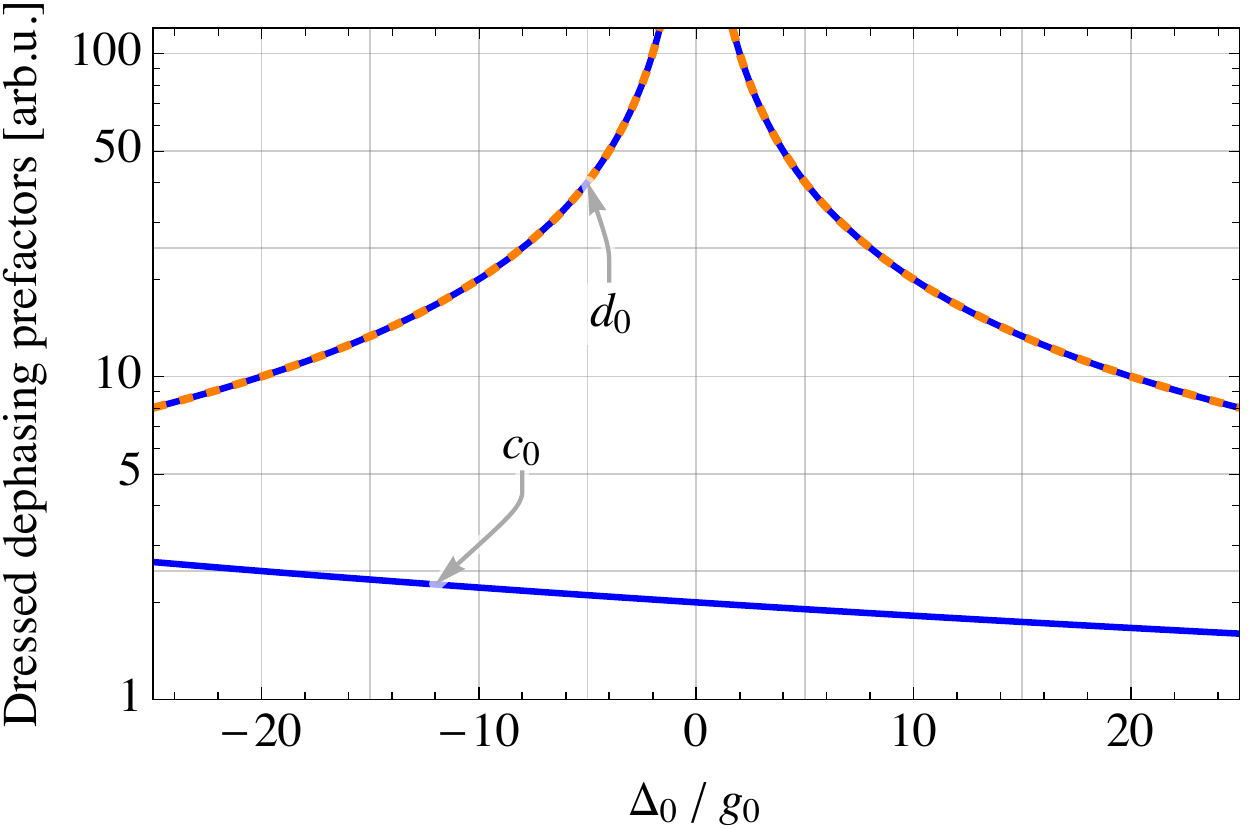}
			\caption{Comparison of the dressed dephasing rate prefactors $d_{0}$ and $c_{0}$, Eqs.~\eqref{eq:DressedPreRabi}, ~\eqref{eq:DressedPreJC} 
				when taking the Rabi (solid blue) or Jaynes-Cummings interaction (dashed orange). 
				The upper curves show $d_{0}$, which is the same for Rabi and Jaynes-Cummings model, 
				while the lower curve shows $c_{0}$, which is exclusive to the Rabi interaction case.
				}
			\label{fig:DressedDephasing}
		\end{center}
	\end{figure}

	The so-called dressed dephasing dissipative corrections have previously been derived 
	in the context of the usual dispersive transformation of the Jaynes-Cummings model~\cite{Boissonneault:PRA:2008, Slichter:PRL:2012}. 
	They describe correlated decay and excitation of qubit and resonator due to coupling to the longitudinal bath of the qubit. 
	In the usual second order perturbation theory this bath only leads to qubit dephasing proportional to its low frequency response.
	Here, due to the resonator-qubit hybridisation, its high frequency components contribute additionally to incoherent conversion of qubit and resonator photons. 
	Using the Keldysh diagrammatic perturbation technique, we reproduce the previously known results, and find additional contributions that are unique to the Rabi model. 
	The total dressed dephasing contribution to the master equation is written as
	\begin{align}
		\mathcal L_{\updownarrow,\pm} \rho 	
			= \sum_k& \gamma_{\downarrow,+}^{(k)} \mathcal D \left[ \sigma_{k,k+1} a^\dagger \right] \rho + \gamma_{\uparrow,-}^{(k)} \mathcal D \left[ \sigma_{k+1,k} a \right] \rho \nn \\
			+& \gamma_{\downarrow,-}^{(k)} \mathcal D \left[ \sigma_{k,k+1} a \right] \rho + \gamma_{\uparrow,+}^{(k)} \mathcal D \left[ \sigma_{k+1,k} a^\dagger \right] \rho \label{eq:DD}\,,
	\end{align}
	with the expressions for the rates
	\begin{align}
		\gamma_{\downarrow,+}^{(k)} &= d_{k} C_{\op Z} \left( \omega_{k+1,k} - \omega_r \right) \,, \quad
		\gamma_{\uparrow,-}^{(k)} = d_{k} C_{\op Z} \left( \omega_r - \omega_{k+1,k} \right) \,,\nn\\
		\gamma_{\downarrow,-}^{(k)} &= c_{k}  C_{\op Z} \left( \omega_r + \omega_{k+1,k} \right) \,, \quad
		\gamma_{\uparrow,+}^{(k)} = c_{k} C_{\op Z} \left( - \omega_{k+1,k} -\omega_r \right) \,,
	\end{align}
	and the dimensionless prefactors for the Rabi model
	\begin{align}
		d_{k,\text{Rabi}} &= \frac{2 g_{k}^2 \:(\delta\omega_{k} - \delta\omega_{k+1})^2}{(\omega_r - \omega_{k+1,k})^2} \,,\nn\\
		c_{k,\text{Rabi}} &= \frac{2 g_{k}^2 \: (\delta\omega_{k} - \delta\omega_{k+1})^2}{(\omega_r + \omega_{k+1,k})^2}\,,\label{eq:DressedPreRabi}
	\end{align}
	and for the Jaynes-Cummings approximation
	\begin{align}
		d_{k,\text{JC}} = d_{k, \text{Rabi}} \,, \quad c_{k,\text{JC}} = 0 \,.\label{eq:DressedPreJC}
	\end{align}
	The first two processes in Eq.~\eqref{eq:DD} correspond to environmentally assisted conversion of photons between the qubit and the resonator, 
	and are equal for both the Rabi and the Jaynes-Cummings model. 
	Exclusive to the Rabi model is the second line in Eq.~\eqref{eq:DD}, corresponding to simultaneous creation or annihilation of a qubit and resonator photon.
	Fig.~\ref{fig:DressedDephasing} shows a comparison of the relative strength of these processes. 
	Note that the rates are proportional to the spectral function of the qubit's longitudinal bath, $C_{{\op Z}}(\omega)$, probed at the sum or difference of qubit and resonator frequencies, 
	$\omega_{k+1,k}\pm\omega_{r}$. 

	\subsubsection{Photon assisted dephasing}
	
\begin{figure}[htbp]
		\begin{center}
			\includegraphics[width=.95\columnwidth]{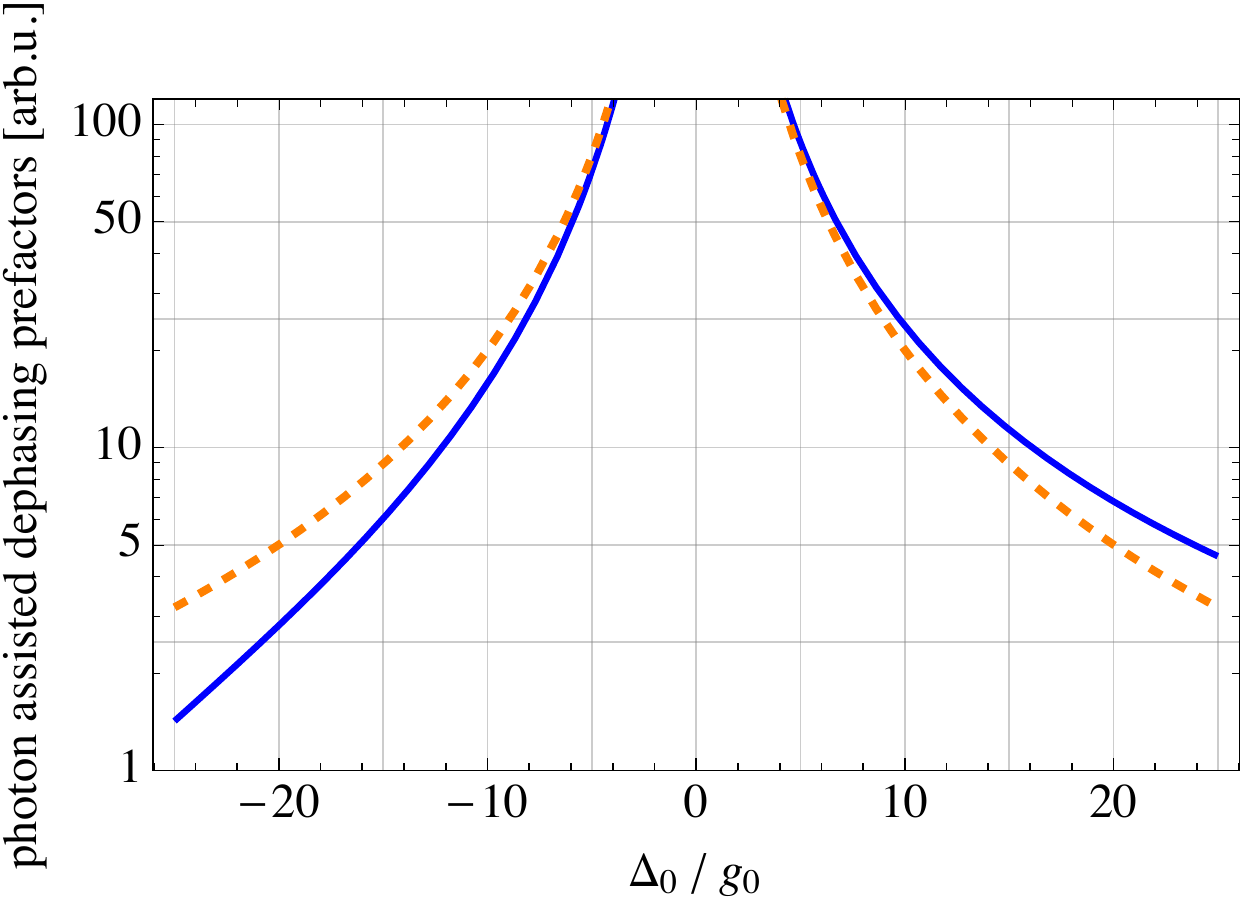}
			\caption{Comparison of the photon assisted dephasing rate prefactors $a_{0}$, Eqs.~\eqref{eq:DephasingAssistedPreRabi},~\eqref{eq:DephasingAssistedPreJC}
				when taking the Rabi (solid blue) or Jaynes-Cummings interaction (dashed orange). 
				The Rabi interaction again leads to a notable asymmetry with respect to zero detuning, in contrast to the Jaynes-Cummings result.}
			\label{fig:DephasingAssisted}
		\end{center}
	\end{figure}

	The final class of process we include at fourth order is typically not considered in the literature, although its contribution to qubit linewidth is likely comparable to Purcell decay. 
	It corresponds to creation or annihilation of a resonator photon together with a simultaneous dephasing process acting on the qubit.
	This has previously been discussed in the context of lasing in double quantum dots~\cite{Mueller:PRA:2016}, where the description of resonator gain and loss was the focus of investigation.
	We write its master equation contribution as
	\begin{align}
		\mathcal L_{\varphi,\pm}\rho 
			= \sum_k \gamma_{\varphi,-}^{(k)} \mathcal D\left[ \sigma_{k,k} a \right] \rho + \gamma_{\varphi,+}^{(k)} \mathcal D\left[ \sigma_{k,k} a^\dagger \right] \rho \,,
		\label{eq:DA}
	\end{align}
	with the rates
	\begin{align}
		\gamma_{\varphi,-}^{(k)} &= a_{k} C_{\op X} \left( \omega_r \right) \,,\quad
		\gamma_{\varphi,+}^{(k)} = a_{k} C_{\op X} \left( -\omega_r \right)
	\end{align}
	and the dimensionless prefactor for the Rabi and Jaynes-Cummings model
	\begin{align}
		a_{k,\text{Rabi}} =& \frac{8g_k^{2} \beta_k^{2} \omega_{k+1,k}^{2}}{(\omega_r^{2} - \omega_{k+1,k}^{2})^2} 
			+ \frac{8g_{k-1}^{2} \beta_{k-1}^{2} \omega_{k,k-1}^{2}}{(\omega_r^{2} - \omega_{k,k-1}^{2})^2} \nn\\
			&-\frac{16 g_{k} g_{k-1} \beta_{k} \beta_{k-1} \omega_{k+1,k}\omega_{k,k-1}}{(\omega_r^{2} - \omega_{k,k-1}^{2})  (\omega_r^{2} - \omega_{k+1,k}^{2})} 
				\label{eq:DephasingAssistedPreRabi} \,,\\
		a_{k,\text{JC}} =& \frac{2 g_k^{2} \beta_k^{2}}{(\omega_r - \omega_{k+1,k})^2} + \frac{2 g_{k-1}^{2} \beta_{k-1}^{2}}{(\omega_r - \omega_{k,k-1})^2} \nn\\
			&- \frac{4 g_{k} g_{k-1} \beta_{k} \beta_{k-1}}{(\omega_r - \omega_{k,k-1}) (\omega_r - \omega_{k+1,k})}
				\label{eq:DephasingAssistedPreJC} \,. 
	\end{align}
	Fig.~\ref{fig:DephasingAssisted} shows a comparison of the rate prefactors for the Rabi and Jaynes-Cummings interaction. 
	As previously, the inclusion of the Rabi interaction terms leads to a pronounced asymmetry in the rates with respect to zero detuning between qubit and resonator, 
	not seen in the Jaynes-Cummings model.
	This dissipative process is proportional to the qubit's transversal bath spectral function, $C_{{\op X}}(\omega)$, probed at the resonator frequency $\omega_{r}$.
	
	\subsubsection{Relevance to the readout problem}
	
	Current prototype implementations of quantum processors all suffer from comparatively poor read-out fidelities. 
	This is at least partly due to the so far imperfect theoretical understanding of the relatively high-powered readout employed in these architectures.
	As an example, Purcell decay of the qubit scales with readout photon number $n$ in the cavity as $\sim 1/n$~\cite{Sete:PRA:2014, Slichter:NJP:2016}, 
	while experimentally it has been shown that the qubit lifetime decreases with increased readout power~\cite{Minev:N:2019}, implying that other processes are dominant in this situation.
	A way out may be offered by recent development of drive induced dissipative processes, 
	which typically arise due to correlated decay processes, like the ones we also calculate here~\cite{Boissonneault:PRA:2009, Boissonneault:PRA:2008, Petrescu:PRB:2020}. 
	In the following we give a short discussion of what the implications of our result are in the strong driving case.
	
	To describe the effect of driving the resonator, we start by applying a displacement transformation to the master equation
	\footnote{
	In the strong driving case, the displacement transformation should actually be applied first, before other perturbative expansions are performed. 
	This would change some of the frequency dependence in the dissipators without changing their qualitative behaviour. 
	For the purpose of simplicity we do not consider this additional complication here. 
	}, 
	$D(\alpha) = \exp\{\alpha a\hc - \alpha\cc a\}$, which transforms resonator operators as 
	\begin{align}
		D\hc a D  = \tilde a + \alpha 
		\label{eq:Displace}
	\end{align}
	with the coherent state amplitude $\alpha$ and the new resonator annihilation operator in the displaced frame $\tilde a$. 
	In this approach, one choses the displacement amplitude $\alpha$ such that the effective resonator variable in the displaced frame, $\tilde a$, now describes an undriven resonator. 
	This is achieved by choosing a dynamical value of 
	$\alpha$ which effectively cancels out the drive terms in the dynamical equations. 
	$\alpha$ then describes the semi-classical dynamics of the photon number in the driven resonator, $n = \abss\alpha$, 
	while the residual photonic quantum fluctuations are still described by the dynamics of $\tilde a$~\cite{Mueller:PRA:2016}.
	Th displacement transformation, \eqn{eq:Displace}, acts on correlated dissipators in a peculiar way. 
	E.g. one of the dressed dephasing dissipators in Eq.~\eqref{eq:DD} transforms as
	\begin{align}
		\mathcal D\left[ D^\dagger \sigma_{k,k+1} a^\dagger D \right] = \mathcal D\left[ \sigma_{k,k+1} \tilde a^\dagger \right] + |\alpha|^2 \mathcal D \left[ \sigma_{k,k+1} \right] + \ldots
		\label{eq:Drive}
	\end{align}
	where the additional terms omitted here will {not contribute to the qubit's dynamics, i.e. will }cancel once a trace over resonator degrees of freedom is performed.
	From the second term in Eq.~\eqref{eq:Drive} we see that {in the driven frame now appears }
	a qubit relaxation channel whose decay rate is proportional to the resonator photon number $n = \abss\alpha$.
	A similar expression holds for all the dressed dephasing terms, 
	which will contribute qubit decay and excitation rates proportional to the photon number. {In the strongly driven case, and reduced to the qubit dynamics, \eqn{eq:DD} then changes to
	\begin{align}
		\mathcal L_{\updownarrow,\pm} \rho \rightarrow \sum_{k} \gamma_{\downarrow}^{(k)} \diss{\sigma_{k, k+1}}\rho + \gamma_{\uparrow}^{(k)} \diss{\sigma_{k+1, k}}\rho
	\end{align}
	with the photon number-dependent rates $\gamma_{\downarrow}^{(k)} = n ( \gamma_{\downarrow,+}^{(k)} +  \gamma_{\downarrow,-}^{(k)})$ and $\gamma_{\uparrow}^{(k)} = n (\gamma_{\uparrow,+}^{(k)} +  \gamma_{\uparrow,-}^{(k)})$.
	Similar corrections have previously been derived in the Jaynes-Cummings approximation~\cite{Boissonneault:PRA:2009}.}
	{Additionally, here we predict a novel photon number dependent contribution to the qubit dephasing rate, from the photon assisted dephasing process, \eqn{eq:DA}, which transforms as
	\begin{align}
		\mathcal L_{\varphi,\pm}\rho \rightarrow \sum_{k} \gamma_{\varphi}^{(k)} \diss{\sigma_{k,k}}\rho
	\end{align} 
	with $\gamma_{\varphi}^{(k)} = n (\gamma_{\varphi,+}^{(k)} + \gamma_{\varphi,-}^{(k)} )$.}	
	For superconducting {transmon} qubits, where the longitudinal bath spectral function is thought to decay as $C_{\op Z}(\omega) \sim 1/\omega$, 
	the major contribution to qubit linewidth here will likely be through {the second process of} photon assisted dephasing.
	A proper comparison to experiments is unfortunately outside the scope of this work at the current time, 
	as experimental investigations are still in the early stages. 
	Additionally, the frequency dependence of the spectral functions of the dissipative baths of both qubit and resonator play a nontrivial role 
	and are currently relatively poorly understood.

\section{Conclusion and Outlook}

	The Jaynes-Cummings model in the dispersive regime is of major importance for todays quantum computing prototypes based on superconducting qubits.
	At its heart it is based on the rotating wave approximation of the underlying Rabi model, and all expression derived from it are subject to the limitations of this approximation. 
	A better analytic understanding of the dispersive parameter regime is fundamental for further improving prototype quantum processors based on superconducting qubits, 
	and will similarly affect the upcoming efforts on large-scale spin qubit devices. 
	
	The analytic equations developed here are based on the Rabi model directly, and allow a more accurate description of quantum systems in the dispersive regime from measured quantities, 
	and thus improved predictions for their behaviour.
	Further, going to higher orders perturbation is straightforward using the Keldysh diagrammatic technique employed here. 
	This may be used in the future to further improve the expressions for the dispersive shifts or to investigate higher order dissipative processes.

	\acknowledgments{I want to thank T.~M.~Stace for help with the early stages of the project, much useful feedback and making me chase up those pesky minus signs, 
	A.~Blais and A.~DiPaolo for suggesting the project in the first place, A.~Petrescu for making me aware of the measurement problem and 
	A.~Blais, J.~H.~Cole, J.~Delaforce, A.~Fedorov, and S.~Filipp for general discussions and feedback on the manuscript. 
	This work was supported by the Australian Research Council Centre of Excellence for Engineered Quantum Systems (EQuS, CE110001013 and CE170100009)
	and the Swiss National Science Fund through NCCR QSIT.}

\bibliography{../DD_KeldyshNotes}

\begin{thebibliography}{27}%
\makeatletter
\providecommand \@ifxundefined [1]{%
 \@ifx{#1\undefined}
}%
\providecommand \@ifnum [1]{%
 \ifnum #1\expandafter \@firstoftwo
 \else \expandafter \@secondoftwo
 \fi
}%
\providecommand \@ifx [1]{%
 \ifx #1\expandafter \@firstoftwo
 \else \expandafter \@secondoftwo
 \fi
}%
\providecommand \natexlab [1]{#1}%
\providecommand \enquote  [1]{``#1''}%
\providecommand \bibnamefont  [1]{#1}%
\providecommand \bibfnamefont [1]{#1}%
\providecommand \citenamefont [1]{#1}%
\providecommand \href@noop [0]{\@secondoftwo}%
\providecommand \href [0]{\begingroup \@sanitize@url \@href}%
\providecommand \@href[1]{\@@startlink{#1}\@@href}%
\providecommand \@@href[1]{\endgroup#1\@@endlink}%
\providecommand \@sanitize@url [0]{\catcode `\\12\catcode `\$12\catcode
  `\&12\catcode `\#12\catcode `\^12\catcode `\_12\catcode `\%12\relax}%
\providecommand \@@startlink[1]{}%
\providecommand \@@endlink[0]{}%
\providecommand \url  [0]{\begingroup\@sanitize@url \@url }%
\providecommand \@url [1]{\endgroup\@href {#1}{\urlprefix }}%
\providecommand \urlprefix  [0]{URL }%
\providecommand \Eprint [0]{\href }%
\providecommand \doibase [0]{http://dx.doi.org/}%
\providecommand \selectlanguage [0]{\@gobble}%
\providecommand \bibinfo  [0]{\@secondoftwo}%
\providecommand \bibfield  [0]{\@secondoftwo}%
\providecommand \translation [1]{[#1]}%
\providecommand \BibitemOpen [0]{}%
\providecommand \bibitemStop [0]{}%
\providecommand \bibitemNoStop [0]{.\EOS\space}%
\providecommand \EOS [0]{\spacefactor3000\relax}%
\providecommand \BibitemShut  [1]{\csname bibitem#1\endcsname}%
\let\auto@bib@innerbib\@empty
\bibitem [{\citenamefont {Walls}\ and\ \citenamefont
  {Milburn}(2008)}]{QuantumOpticsBook}%
  \BibitemOpen
  \bibfield  {author} {\bibinfo {author} {\bibfnamefont {D.~F.}\ \bibnamefont
  {Walls}}\ and\ \bibinfo {author} {\bibfnamefont {G.~J.}\ \bibnamefont
  {Milburn}},\ }\href {\doibase 10.1007/978-3-540-28574-8} {\emph {\bibinfo
  {title} {Quantum Optics}}}\ (\bibinfo  {publisher} {Springer, Berlin},\
  \bibinfo {year} {2008})\BibitemShut {NoStop}%
\bibitem [{\citenamefont {Jaynes}\ and\ \citenamefont
  {Cummings}(1963)}]{JaynesCummings}%
  \BibitemOpen
  \bibfield  {author} {\bibinfo {author} {\bibfnamefont {E.~T.}\ \bibnamefont
  {Jaynes}}\ and\ \bibinfo {author} {\bibfnamefont {F.~W.}\ \bibnamefont
  {Cummings}},\ }\bibfield  {title} {\enquote {\bibinfo {title} {Comparison of
  quantum and semiclassical radiation theories with application to the beam
  maser},}\ }\href@noop {} {\bibfield  {journal} {\bibinfo  {journal}
  {Proceedings of the IEEE}\ }\textbf {\bibinfo {volume} {51}},\ \bibinfo
  {pages} {89 -- 109} (\bibinfo {year} {1963})}\BibitemShut {NoStop}%
\bibitem [{\citenamefont {Blais}\ \emph {et~al.}(2004)\citenamefont {Blais},
  \citenamefont {Huang}, \citenamefont {Wallraff}, \citenamefont {Girvin},\
  and\ \citenamefont {Schoelkopf}}]{Blais:PRA:2004}%
  \BibitemOpen
  \bibfield  {author} {\bibinfo {author} {\bibfnamefont {Alexandre}\
  \bibnamefont {Blais}}, \bibinfo {author} {\bibfnamefont {Ren-Shou}\
  \bibnamefont {Huang}}, \bibinfo {author} {\bibfnamefont {Andreas}\
  \bibnamefont {Wallraff}}, \bibinfo {author} {\bibfnamefont {Steven~M}\
  \bibnamefont {Girvin}}, \ and\ \bibinfo {author} {\bibfnamefont {Robert~J}\
  \bibnamefont {Schoelkopf}},\ }\bibfield  {title} {\enquote {\bibinfo {title}
  {{Cavity quantum electrodynamics for superconducting electrical circuits: An
  architecture for quantum computation}},}\ }\href {\doibase
  10.1103/PhysRevA.69.062320} {\bibfield  {journal} {\bibinfo  {journal} {Phys.
  Rev. A}\ }\textbf {\bibinfo {volume} {69}},\ \bibinfo {pages} {062320}
  (\bibinfo {year} {2004})}\BibitemShut {NoStop}%
\bibitem [{\citenamefont {Rabi}(1937)}]{RabiModel}%
  \BibitemOpen
  \bibfield  {author} {\bibinfo {author} {\bibfnamefont {I.~I.}\ \bibnamefont
  {Rabi}},\ }\bibfield  {title} {\enquote {\bibinfo {title} {Space quantization
  in a gyrating magnetic field},}\ }\href {\doibase 10.1103/PhysRev.51.652}
  {\bibfield  {journal} {\bibinfo  {journal} {Phys. Rev.}\ }\textbf {\bibinfo
  {volume} {51}},\ \bibinfo {pages} {652} (\bibinfo {year} {1937})}\BibitemShut
  {NoStop}%
\bibitem [{\citenamefont {Forn-Diaz}\ \emph {et~al.}(2016)\citenamefont
  {Forn-Diaz}, \citenamefont {Garc{\'\i}a-Ripoll}, \citenamefont {Peroprade},
  \citenamefont {Orgiazzi}, \citenamefont {Yurtalan}, \citenamefont
  {Belyansky}, \citenamefont {Wislon},\ and\ \citenamefont
  {Lupascu}}]{FornDiaz:NP:2016}%
  \BibitemOpen
  \bibfield  {author} {\bibinfo {author} {\bibfnamefont {P.}~\bibnamefont
  {Forn-Diaz}}, \bibinfo {author} {\bibfnamefont {J.~J.}\ \bibnamefont
  {Garc{\'\i}a-Ripoll}}, \bibinfo {author} {\bibfnamefont {B.}~\bibnamefont
  {Peroprade}}, \bibinfo {author} {\bibfnamefont {J.-L.}\ \bibnamefont
  {Orgiazzi}}, \bibinfo {author} {\bibfnamefont {M.~A.}\ \bibnamefont
  {Yurtalan}}, \bibinfo {author} {\bibfnamefont {R.}~\bibnamefont {Belyansky}},
  \bibinfo {author} {\bibfnamefont {C.~M.}\ \bibnamefont {Wislon}}, \ and\
  \bibinfo {author} {\bibfnamefont {A.}~\bibnamefont {Lupascu}},\ }\bibfield
  {title} {\enquote {\bibinfo {title} {Ultrastrong coupling of a single
  artificial atom to an electromagnetic continuum in the nonperturbative
  regime},}\ }\href {\doibase 10.1038/nphys3905} {\bibfield  {journal}
  {\bibinfo  {journal} {Nature Physics}\ }\textbf {\bibinfo {volume} {13}},\
  \bibinfo {pages} {39--43} (\bibinfo {year} {2016})}\BibitemShut {NoStop}%
\bibitem [{\citenamefont {Kockum}\ \emph {et~al.}(2019)\citenamefont {Kockum},
  \citenamefont {Miranowicz}, \citenamefont {De~Liberato}, \citenamefont
  {Savasta},\ and\ \citenamefont {Nori}}]{Kockum:NR:2019}%
  \BibitemOpen
  \bibfield  {author} {\bibinfo {author} {\bibfnamefont {A.~F.}\ \bibnamefont
  {Kockum}}, \bibinfo {author} {\bibfnamefont {A.}~\bibnamefont {Miranowicz}},
  \bibinfo {author} {\bibfnamefont {S.}~\bibnamefont {De~Liberato}}, \bibinfo
  {author} {\bibfnamefont {S.}~\bibnamefont {Savasta}}, \ and\ \bibinfo
  {author} {\bibfnamefont {F.}~\bibnamefont {Nori}},\ }\bibfield  {title}
  {\enquote {\bibinfo {title} {Ultrastrong coupling between light and
  matter},}\ }\href {\doibase 10.1038/s42254-018-0006-2} {\bibfield  {journal}
  {\bibinfo  {journal} {Nature Reviews Physics}\ }\textbf {\bibinfo {volume}
  {1}},\ \bibinfo {pages} {19--40} (\bibinfo {year} {2019})}\BibitemShut
  {NoStop}%
\bibitem [{\citenamefont {Chow}\ \emph {et~al.}(2014)\citenamefont {Chow},
  \citenamefont {Gambetta}, \citenamefont {Magesan}, \citenamefont {Abraham},
  \citenamefont {Cross}, \citenamefont {Johnson}, \citenamefont {Masluk},
  \citenamefont {Ryan}, \citenamefont {Smolin}, \citenamefont {Srinivasan},\
  and\ \citenamefont {Steffen}}]{Chow:NC:2014}%
  \BibitemOpen
  \bibfield  {author} {\bibinfo {author} {\bibfnamefont {Jerry~M}\ \bibnamefont
  {Chow}}, \bibinfo {author} {\bibfnamefont {Jay~M}\ \bibnamefont {Gambetta}},
  \bibinfo {author} {\bibfnamefont {Easwar}\ \bibnamefont {Magesan}}, \bibinfo
  {author} {\bibfnamefont {David~W}\ \bibnamefont {Abraham}}, \bibinfo {author}
  {\bibfnamefont {Andrew~W}\ \bibnamefont {Cross}}, \bibinfo {author}
  {\bibfnamefont {Blake~R}\ \bibnamefont {Johnson}}, \bibinfo {author}
  {\bibfnamefont {Nicholas~A}\ \bibnamefont {Masluk}}, \bibinfo {author}
  {\bibfnamefont {Colm~A}\ \bibnamefont {Ryan}}, \bibinfo {author}
  {\bibfnamefont {John}\ \bibnamefont {Smolin}}, \bibinfo {author}
  {\bibfnamefont {Srikanth~J}\ \bibnamefont {Srinivasan}}, \ and\ \bibinfo
  {author} {\bibfnamefont {Matthias}\ \bibnamefont {Steffen}},\ }\bibfield
  {title} {\enquote {\bibinfo {title} {Implementing a strand of a scalable
  fault-tolerant quantum computing fabric},}\ }\href {\doibase
  10.1038/ncomms5015} {\bibfield  {journal} {\bibinfo  {journal} {Nature
  Communications}\ }\textbf {\bibinfo {volume} {5}},\ \bibinfo {pages} {4015}
  (\bibinfo {year} {2014})}\BibitemShut {NoStop}%
\bibitem [{\citenamefont {Barends}\ \emph {et~al.}(2016)\citenamefont
  {Barends}, \citenamefont {Shabani}, \citenamefont {Lamata}, \citenamefont
  {Kelly}, \citenamefont {Mezzacapo}, \citenamefont {Las~Heras}, \citenamefont
  {Babbush}, \citenamefont {Fowler}, \citenamefont {Campbell}, \citenamefont
  {Chen}, \citenamefont {Chen}, \citenamefont {Chiaro}, \citenamefont
  {Dunsworth}, \citenamefont {Jeffrey}, \citenamefont {Lucero}, \citenamefont
  {Megrant}, \citenamefont {Mutus}, \citenamefont {Neeley}, \citenamefont
  {Neill}, \citenamefont {O'Malley}, \citenamefont {Quintana}, \citenamefont
  {Roushan}, \citenamefont {Sank}, \citenamefont {Vainsencher}, \citenamefont
  {Wenner}, \citenamefont {White}, \citenamefont {Solano}, \citenamefont
  {Neven},\ and\ \citenamefont {Martinis}}]{Barends:N:2016}%
  \BibitemOpen
  \bibfield  {author} {\bibinfo {author} {\bibfnamefont {Rami}\ \bibnamefont
  {Barends}}, \bibinfo {author} {\bibfnamefont {A}~\bibnamefont {Shabani}},
  \bibinfo {author} {\bibfnamefont {Lucas}\ \bibnamefont {Lamata}}, \bibinfo
  {author} {\bibfnamefont {J}~\bibnamefont {Kelly}}, \bibinfo {author}
  {\bibfnamefont {Antonio}\ \bibnamefont {Mezzacapo}}, \bibinfo {author}
  {\bibfnamefont {U}~\bibnamefont {Las~Heras}}, \bibinfo {author}
  {\bibfnamefont {R}~\bibnamefont {Babbush}}, \bibinfo {author} {\bibfnamefont
  {Austin~G}\ \bibnamefont {Fowler}}, \bibinfo {author} {\bibfnamefont
  {B}~\bibnamefont {Campbell}}, \bibinfo {author} {\bibfnamefont
  {Yu}~\bibnamefont {Chen}}, \bibinfo {author} {\bibfnamefont {Z}~\bibnamefont
  {Chen}}, \bibinfo {author} {\bibfnamefont {B}~\bibnamefont {Chiaro}},
  \bibinfo {author} {\bibfnamefont {A}~\bibnamefont {Dunsworth}}, \bibinfo
  {author} {\bibfnamefont {E}~\bibnamefont {Jeffrey}}, \bibinfo {author}
  {\bibfnamefont {Erik}\ \bibnamefont {Lucero}}, \bibinfo {author}
  {\bibfnamefont {A}~\bibnamefont {Megrant}}, \bibinfo {author} {\bibfnamefont
  {J~Y}\ \bibnamefont {Mutus}}, \bibinfo {author} {\bibfnamefont {Matthew}\
  \bibnamefont {Neeley}}, \bibinfo {author} {\bibfnamefont {Charles}\
  \bibnamefont {Neill}}, \bibinfo {author} {\bibfnamefont {P~J~J}\ \bibnamefont
  {O'Malley}}, \bibinfo {author} {\bibfnamefont {Chris}\ \bibnamefont
  {Quintana}}, \bibinfo {author} {\bibfnamefont {Pedram}\ \bibnamefont
  {Roushan}}, \bibinfo {author} {\bibfnamefont {D}~\bibnamefont {Sank}},
  \bibinfo {author} {\bibfnamefont {Amit}\ \bibnamefont {Vainsencher}},
  \bibinfo {author} {\bibfnamefont {James}\ \bibnamefont {Wenner}}, \bibinfo
  {author} {\bibfnamefont {T.~C.}\ \bibnamefont {White}}, \bibinfo {author}
  {\bibfnamefont {Enrique}\ \bibnamefont {Solano}}, \bibinfo {author}
  {\bibfnamefont {H}~\bibnamefont {Neven}}, \ and\ \bibinfo {author}
  {\bibfnamefont {John~M}\ \bibnamefont {Martinis}},\ }\bibfield  {title}
  {\enquote {\bibinfo {title} {Digitized adiabatic quantum computing with a
  superconducting circuit},}\ }\href {\doibase 10.1038/nature17658} {\bibfield
  {journal} {\bibinfo  {journal} {Nature}\ }\textbf {\bibinfo {volume} {534}},\
  \bibinfo {pages} {222--226} (\bibinfo {year} {2016})}\BibitemShut {NoStop}%
\bibitem [{\citenamefont {Roushan}\ \emph {et~al.}(2017)\citenamefont
  {Roushan}, \citenamefont {Neill}, \citenamefont {Tangpanitanon},
  \citenamefont {Bastidas}, \citenamefont {Megrant}, \citenamefont {Barends},
  \citenamefont {Chen}, \citenamefont {Chen}, \citenamefont {Chiaro},
  \citenamefont {Dunsworth}, \citenamefont {Fowler}, \citenamefont {Foxen},
  \citenamefont {Giustina}, \citenamefont {Jeffrey}, \citenamefont {Kelly},
  \citenamefont {Lucero}, \citenamefont {Mutus}, \citenamefont {Neeley},
  \citenamefont {Quintana}, \citenamefont {Sank}, \citenamefont {Vainsencher},
  \citenamefont {Wenner}, \citenamefont {White}, \citenamefont {Neven},
  \citenamefont {Angelakis},\ and\ \citenamefont {Martinis}}]{Roushan:S:2017}%
  \BibitemOpen
  \bibfield  {author} {\bibinfo {author} {\bibfnamefont {Pedram}\ \bibnamefont
  {Roushan}}, \bibinfo {author} {\bibfnamefont {Charles}\ \bibnamefont
  {Neill}}, \bibinfo {author} {\bibfnamefont {J}~\bibnamefont {Tangpanitanon}},
  \bibinfo {author} {\bibfnamefont {Victor~M}\ \bibnamefont {Bastidas}},
  \bibinfo {author} {\bibfnamefont {A}~\bibnamefont {Megrant}}, \bibinfo
  {author} {\bibfnamefont {Rami}\ \bibnamefont {Barends}}, \bibinfo {author}
  {\bibfnamefont {Yu}~\bibnamefont {Chen}}, \bibinfo {author} {\bibfnamefont
  {Z}~\bibnamefont {Chen}}, \bibinfo {author} {\bibfnamefont {B}~\bibnamefont
  {Chiaro}}, \bibinfo {author} {\bibfnamefont {A}~\bibnamefont {Dunsworth}},
  \bibinfo {author} {\bibfnamefont {Austin~G}\ \bibnamefont {Fowler}}, \bibinfo
  {author} {\bibfnamefont {B}~\bibnamefont {Foxen}}, \bibinfo {author}
  {\bibfnamefont {Marissa}\ \bibnamefont {Giustina}}, \bibinfo {author}
  {\bibfnamefont {E}~\bibnamefont {Jeffrey}}, \bibinfo {author} {\bibfnamefont
  {J}~\bibnamefont {Kelly}}, \bibinfo {author} {\bibfnamefont {Erik}\
  \bibnamefont {Lucero}}, \bibinfo {author} {\bibfnamefont {J~Y}\ \bibnamefont
  {Mutus}}, \bibinfo {author} {\bibfnamefont {Matthew}\ \bibnamefont {Neeley}},
  \bibinfo {author} {\bibfnamefont {Chris}\ \bibnamefont {Quintana}}, \bibinfo
  {author} {\bibfnamefont {D}~\bibnamefont {Sank}}, \bibinfo {author}
  {\bibfnamefont {Amit}\ \bibnamefont {Vainsencher}}, \bibinfo {author}
  {\bibfnamefont {James}\ \bibnamefont {Wenner}}, \bibinfo {author}
  {\bibfnamefont {T.~C.}\ \bibnamefont {White}}, \bibinfo {author}
  {\bibfnamefont {H}~\bibnamefont {Neven}}, \bibinfo {author} {\bibfnamefont
  {Dimitris}\ \bibnamefont {Angelakis}}, \ and\ \bibinfo {author}
  {\bibfnamefont {John~M}\ \bibnamefont {Martinis}},\ }\bibfield  {title}
  {\enquote {\bibinfo {title} {Spectroscopic signatures of localization with
  interacting photons in superconducting qubits.}}\ }\href {\doibase
  10.1126/science.aao1401} {\bibfield  {journal} {\bibinfo  {journal}
  {Science}\ }\textbf {\bibinfo {volume} {358}},\ \bibinfo {pages} {1175--1179}
  (\bibinfo {year} {2017})}\BibitemShut {NoStop}%
\bibitem [{\citenamefont {Kandala}\ \emph {et~al.}(2019)\citenamefont
  {Kandala}, \citenamefont {Temme}, \citenamefont {C{\'o}rcoles}, \citenamefont
  {Mezzacapo}, \citenamefont {Chow},\ and\ \citenamefont
  {Gambetta}}]{Kandala:N:2019}%
  \BibitemOpen
  \bibfield  {author} {\bibinfo {author} {\bibfnamefont {A.}~\bibnamefont
  {Kandala}}, \bibinfo {author} {\bibfnamefont {K.}~\bibnamefont {Temme}},
  \bibinfo {author} {\bibfnamefont {A.~D.}\ \bibnamefont {C{\'o}rcoles}},
  \bibinfo {author} {\bibfnamefont {A.}~\bibnamefont {Mezzacapo}}, \bibinfo
  {author} {\bibfnamefont {J.~M.}\ \bibnamefont {Chow}}, \ and\ \bibinfo
  {author} {\bibfnamefont {J.~M.}\ \bibnamefont {Gambetta}},\ }\bibfield
  {title} {\enquote {\bibinfo {title} {Error mitigation extends the
  computational reach of a noisy quantum processor},}\ }\href {\doibase
  10.1038/s41586-019-1040-7} {\bibfield  {journal} {\bibinfo  {journal}
  {Nature}\ }\textbf {\bibinfo {volume} {567}},\ \bibinfo {pages} {491--495}
  (\bibinfo {year} {2019})}\BibitemShut {NoStop}%
\bibitem [{\citenamefont {Krantz}\ \emph {et~al.}(2019)\citenamefont {Krantz},
  \citenamefont {Kjaergaard}, \citenamefont {Yan}, \citenamefont {Orlando},
  \citenamefont {Gustavsson},\ and\ \citenamefont {Oliver}}]{QuantumEngineer}%
  \BibitemOpen
  \bibfield  {author} {\bibinfo {author} {\bibfnamefont {P.}~\bibnamefont
  {Krantz}}, \bibinfo {author} {\bibfnamefont {M.}~\bibnamefont {Kjaergaard}},
  \bibinfo {author} {\bibfnamefont {F.}~\bibnamefont {Yan}}, \bibinfo {author}
  {\bibfnamefont {T.~P.}\ \bibnamefont {Orlando}}, \bibinfo {author}
  {\bibfnamefont {S.}~\bibnamefont {Gustavsson}}, \ and\ \bibinfo {author}
  {\bibfnamefont {W.~D.}\ \bibnamefont {Oliver}},\ }\bibfield  {title}
  {\enquote {\bibinfo {title} {A quantum engineer's guide to superconducting
  qubits},}\ }\href {\doibase 10.1063/1.5089550} {\bibfield  {journal}
  {\bibinfo  {journal} {Applied Physics Review}\ }\textbf {\bibinfo {volume}
  {6}},\ \bibinfo {pages} {021318} (\bibinfo {year} {2019})}\BibitemShut
  {NoStop}%
\bibitem [{\citenamefont {Petersson}\ \emph {et~al.}(2012)\citenamefont
  {Petersson}, \citenamefont {McFaul}, \citenamefont {Schroer}, \citenamefont
  {Jung}, \citenamefont {Taylor}, \citenamefont {Houck},\ and\ \citenamefont
  {Petta}}]{Petersson:N:2012}%
  \BibitemOpen
  \bibfield  {author} {\bibinfo {author} {\bibfnamefont {K~D}\ \bibnamefont
  {Petersson}}, \bibinfo {author} {\bibfnamefont {L.~W.}\ \bibnamefont
  {McFaul}}, \bibinfo {author} {\bibfnamefont {M.~D.}\ \bibnamefont {Schroer}},
  \bibinfo {author} {\bibfnamefont {M.}~\bibnamefont {Jung}}, \bibinfo {author}
  {\bibfnamefont {Jacob~M}\ \bibnamefont {Taylor}}, \bibinfo {author}
  {\bibfnamefont {Andrew~A}\ \bibnamefont {Houck}}, \ and\ \bibinfo {author}
  {\bibfnamefont {J~R}\ \bibnamefont {Petta}},\ }\bibfield  {title} {\enquote
  {\bibinfo {title} {Circuit quantum electrodynamics with a spin qubit},}\
  }\href {\doibase 10.1038/nature11559} {\bibfield  {journal} {\bibinfo
  {journal} {Nature}\ }\textbf {\bibinfo {volume} {490}},\ \bibinfo {pages}
  {380--383} (\bibinfo {year} {2012})}\BibitemShut {NoStop}%
\bibitem [{\citenamefont {van Woerkom}\ \emph {et~al.}(2018)\citenamefont {van
  Woerkom}, \citenamefont {Scarlino}, \citenamefont {Ungerer}, \citenamefont
  {M{\"u}ller}, \citenamefont {Koski}, \citenamefont {Landig}, \citenamefont
  {Reichl}, \citenamefont {Wegscheider}, \citenamefont {Ihn}, \citenamefont
  {Ensslin},\ and\ \citenamefont {Wallraff}}]{vanWoerkom:PRX:2018}%
  \BibitemOpen
  \bibfield  {author} {\bibinfo {author} {\bibfnamefont {David~J}\ \bibnamefont
  {van Woerkom}}, \bibinfo {author} {\bibfnamefont {Pasquale}\ \bibnamefont
  {Scarlino}}, \bibinfo {author} {\bibfnamefont {Jann~H}\ \bibnamefont
  {Ungerer}}, \bibinfo {author} {\bibfnamefont {Clemens}\ \bibnamefont
  {M{\"u}ller}}, \bibinfo {author} {\bibfnamefont {Jonne~V}\ \bibnamefont
  {Koski}}, \bibinfo {author} {\bibfnamefont {Andreas~J}\ \bibnamefont
  {Landig}}, \bibinfo {author} {\bibfnamefont {Christian}\ \bibnamefont
  {Reichl}}, \bibinfo {author} {\bibfnamefont {Werner}\ \bibnamefont
  {Wegscheider}}, \bibinfo {author} {\bibfnamefont {Thomas~Markus}\
  \bibnamefont {Ihn}}, \bibinfo {author} {\bibfnamefont {Klaus}\ \bibnamefont
  {Ensslin}}, \ and\ \bibinfo {author} {\bibfnamefont {Andreas}\ \bibnamefont
  {Wallraff}},\ }\bibfield  {title} {\enquote {\bibinfo {title} {Microwave
  photon-mediated interactions between semiconductor qubits},}\ }\href
  {\doibase 10.1103/PhysRevX.8.041018} {\bibfield  {journal} {\bibinfo
  {journal} {Physical Review X}\ }\textbf {\bibinfo {volume} {8}},\ \bibinfo
  {pages} {041018} (\bibinfo {year} {2018})}\BibitemShut {NoStop}%
\bibitem [{\citenamefont {Landig}\ \emph {et~al.}(2019)\citenamefont {Landig},
  \citenamefont {Koski}, \citenamefont {Scarlino}, \citenamefont {M{\"u}ller},
  \citenamefont {Abadillo-Uriel}, \citenamefont {Kratochwil}, \citenamefont
  {Reichl}, \citenamefont {Wegscheider}, \citenamefont {Coppersmith},
  \citenamefont {Friesen}, \citenamefont {Wallraff}, \citenamefont {Ihn},\ and\
  \citenamefont {Ensslin}}]{Landig:NC:2019}%
  \BibitemOpen
  \bibfield  {author} {\bibinfo {author} {\bibfnamefont {Andreas~J}\
  \bibnamefont {Landig}}, \bibinfo {author} {\bibfnamefont {Jonne~V}\
  \bibnamefont {Koski}}, \bibinfo {author} {\bibfnamefont {Pasquale}\
  \bibnamefont {Scarlino}}, \bibinfo {author} {\bibfnamefont {Clemens}\
  \bibnamefont {M{\"u}ller}}, \bibinfo {author} {\bibfnamefont {Jos{\'e}~C}\
  \bibnamefont {Abadillo-Uriel}}, \bibinfo {author} {\bibfnamefont
  {B}~\bibnamefont {Kratochwil}}, \bibinfo {author} {\bibfnamefont {Christian}\
  \bibnamefont {Reichl}}, \bibinfo {author} {\bibfnamefont {W}~\bibnamefont
  {Wegscheider}}, \bibinfo {author} {\bibfnamefont {S~N}\ \bibnamefont
  {Coppersmith}}, \bibinfo {author} {\bibfnamefont {Mark}\ \bibnamefont
  {Friesen}}, \bibinfo {author} {\bibfnamefont {Andreas}\ \bibnamefont
  {Wallraff}}, \bibinfo {author} {\bibfnamefont {Thomas~Markus}\ \bibnamefont
  {Ihn}}, \ and\ \bibinfo {author} {\bibfnamefont {Klaus}\ \bibnamefont
  {Ensslin}},\ }\bibfield  {title} {\enquote {\bibinfo {title}
  {Virtual-photon-mediated spin-qubit-transmon coupling},}\ }\href {\doibase
  10.1038/s41467-019-13000-z} {\bibfield  {journal} {\bibinfo  {journal}
  {Nature Communications}\ }\textbf {\bibinfo {volume} {10}},\ \bibinfo {pages}
  {5037} (\bibinfo {year} {2019})}\BibitemShut {NoStop}%
\bibitem [{\citenamefont {Boissonneault}\ \emph {et~al.}(2008)\citenamefont
  {Boissonneault}, \citenamefont {Gambetta},\ and\ \citenamefont
  {Blais}}]{Boissonneault:PRA:2008}%
  \BibitemOpen
  \bibfield  {author} {\bibinfo {author} {\bibfnamefont {Maxime}\ \bibnamefont
  {Boissonneault}}, \bibinfo {author} {\bibfnamefont {Jay~M}\ \bibnamefont
  {Gambetta}}, \ and\ \bibinfo {author} {\bibfnamefont {Alexandre}\
  \bibnamefont {Blais}},\ }\bibfield  {title} {\enquote {\bibinfo {title}
  {Nonlinear dispersive regime of cavity {QED}: The dressed dephasing model},}\
  }\href {\doibase 10.1103/PhysRevA.77.060305} {\bibfield  {journal} {\bibinfo
  {journal} {Physical Review A}\ }\textbf {\bibinfo {volume} {77}},\ \bibinfo
  {pages} {060305(R)} (\bibinfo {year} {2008})}\BibitemShut {NoStop}%
\bibitem [{\citenamefont {Boissonneault}\ \emph {et~al.}(2009)\citenamefont
  {Boissonneault}, \citenamefont {Gambetta},\ and\ \citenamefont
  {Blais}}]{Boissonneault:PRA:2009}%
  \BibitemOpen
  \bibfield  {author} {\bibinfo {author} {\bibfnamefont {Maxime}\ \bibnamefont
  {Boissonneault}}, \bibinfo {author} {\bibfnamefont {Jay~M}\ \bibnamefont
  {Gambetta}}, \ and\ \bibinfo {author} {\bibfnamefont {Alexandre}\
  \bibnamefont {Blais}},\ }\bibfield  {title} {\enquote {\bibinfo {title}
  {Dispersive regime of circuit {QED}: Photon-dependent qubit dephasing and
  relaxation rates},}\ }\href {\doibase 10.1103/PhysRevA.79.013819} {\bibfield
  {journal} {\bibinfo  {journal} {Physical Review A}\ }\textbf {\bibinfo
  {volume} {79}},\ \bibinfo {pages} {013819} (\bibinfo {year}
  {2009})}\BibitemShut {NoStop}%
\bibitem [{\citenamefont {M{\"u}ller}\ and\ \citenamefont
  {Stace}(2017)}]{Mueller:PRA:2016}%
  \BibitemOpen
  \bibfield  {author} {\bibinfo {author} {\bibfnamefont {Clemens}\ \bibnamefont
  {M{\"u}ller}}\ and\ \bibinfo {author} {\bibfnamefont {Thomas~M}\ \bibnamefont
  {Stace}},\ }\bibfield  {title} {\enquote {\bibinfo {title} {{Deriving
  Lindblad master equations with Keldysh diagrams: Correlated gain and loss in
  higher order perturbation theory}},}\ }\href {\doibase
  10.1103/PhysRevA.95.013847} {\bibfield  {journal} {\bibinfo  {journal} {Phys.
  Rev. A}\ }\textbf {\bibinfo {volume} {95}},\ \bibinfo {pages} {013847}
  (\bibinfo {year} {2017})}\BibitemShut {NoStop}%
\bibitem [{\citenamefont {Stace}\ \emph {et~al.}(2013)\citenamefont {Stace},
  \citenamefont {Doherty},\ and\ \citenamefont {Reilly}}]{Stace:PRL:2013}%
  \BibitemOpen
  \bibfield  {author} {\bibinfo {author} {\bibfnamefont {Thomas~M}\
  \bibnamefont {Stace}}, \bibinfo {author} {\bibfnamefont {Andrew~C}\
  \bibnamefont {Doherty}}, \ and\ \bibinfo {author} {\bibfnamefont {David~J}\
  \bibnamefont {Reilly}},\ }\bibfield  {title} {\enquote {\bibinfo {title}
  {Dynamical steady states in driven quantum systems},}\ }\href {\doibase
  10.1103/PhysRevLett.111.180602} {\bibfield  {journal} {\bibinfo  {journal}
  {Physical Review Letters}\ }\textbf {\bibinfo {volume} {111}},\ \bibinfo
  {pages} {180602} (\bibinfo {year} {2013})}\BibitemShut {NoStop}%
\bibitem [{\citenamefont {Forn-Diaz}\ \emph {et~al.}(2010)\citenamefont
  {Forn-Diaz}, \citenamefont {Lisenfeld}, \citenamefont {Marcos}, \citenamefont
  {Garcia-Ripoll}, \citenamefont {Solano}, \citenamefont {Harmans},\ and\
  \citenamefont {Mooij}}]{FornDiaz:PRL:2010}%
  \BibitemOpen
  \bibfield  {author} {\bibinfo {author} {\bibfnamefont {P.}~\bibnamefont
  {Forn-Diaz}}, \bibinfo {author} {\bibfnamefont {J{\"u}rgen}\ \bibnamefont
  {Lisenfeld}}, \bibinfo {author} {\bibfnamefont {David}\ \bibnamefont
  {Marcos}}, \bibinfo {author} {\bibfnamefont {Juan~Jos{\'e}}\ \bibnamefont
  {Garcia-Ripoll}}, \bibinfo {author} {\bibfnamefont {Enrique}\ \bibnamefont
  {Solano}}, \bibinfo {author} {\bibfnamefont {C~J P~M}\ \bibnamefont
  {Harmans}}, \ and\ \bibinfo {author} {\bibfnamefont {J~E}\ \bibnamefont
  {Mooij}},\ }\bibfield  {title} {\enquote {\bibinfo {title} {{Observation of
  the Bloch-Siegert Shift in a Qubit-Oscillator System in the Ultrastrong
  Coupling Regime}},}\ }\href {\doibase 10.1103/PhysRevLett.105.237001}
  {\bibfield  {journal} {\bibinfo  {journal} {Physical Review Letters}\
  }\textbf {\bibinfo {volume} {105}},\ \bibinfo {pages} {237001} (\bibinfo
  {year} {2010})}\BibitemShut {NoStop}%
\bibitem [{\citenamefont {Zueco}\ \emph {et~al.}(2009)\citenamefont {Zueco},
  \citenamefont {Reuther}, \citenamefont {Kohler},\ and\ \citenamefont
  {H{\"a}nggi}}]{Zueco:PRA:2009}%
  \BibitemOpen
  \bibfield  {author} {\bibinfo {author} {\bibfnamefont {David}\ \bibnamefont
  {Zueco}}, \bibinfo {author} {\bibfnamefont {Georg~M}\ \bibnamefont
  {Reuther}}, \bibinfo {author} {\bibfnamefont {Sigmund}\ \bibnamefont
  {Kohler}}, \ and\ \bibinfo {author} {\bibfnamefont {Peter}\ \bibnamefont
  {H{\"a}nggi}},\ }\bibfield  {title} {\enquote {\bibinfo {title}
  {Qubit-oscillator dynamics in the dispersive regime: Analytical theory beyond
  the rotating-wave approximation},}\ }\href {\doibase
  10.1103/PhysRevA.80.033846} {\bibfield  {journal} {\bibinfo  {journal}
  {Physical Review A}\ }\textbf {\bibinfo {volume} {80}},\ \bibinfo {pages}
  {033846} (\bibinfo {year} {2009})}\BibitemShut {NoStop}%
\bibitem [{\citenamefont {Zhu}\ \emph {et~al.}(2013)\citenamefont {Zhu},
  \citenamefont {Schmidt},\ and\ \citenamefont {Koch}}]{Zhu:NJP:2013}%
  \BibitemOpen
  \bibfield  {author} {\bibinfo {author} {\bibfnamefont {Guanyu}\ \bibnamefont
  {Zhu}}, \bibinfo {author} {\bibfnamefont {Sebastian}\ \bibnamefont
  {Schmidt}}, \ and\ \bibinfo {author} {\bibfnamefont {Jens}\ \bibnamefont
  {Koch}},\ }\bibfield  {title} {\enquote {\bibinfo {title} {Dispersive regime
  of the jaynes--cummings and rabi lattice},}\ }\href {\doibase
  10.1088/1367-2630/15/11/115002} {\bibfield  {journal} {\bibinfo  {journal}
  {New Journal of Physics}\ }\textbf {\bibinfo {volume} {15}},\ \bibinfo
  {pages} {115002} (\bibinfo {year} {2013})}\BibitemShut {NoStop}%
\bibitem [{\citenamefont {Slichter}\ \emph {et~al.}(2012)\citenamefont
  {Slichter}, \citenamefont {Vijay}, \citenamefont {Weber}, \citenamefont
  {Boutin}, \citenamefont {Boissonneault}, \citenamefont {Gambetta},
  \citenamefont {Blais},\ and\ \citenamefont {Siddiqi}}]{Slichter:PRL:2012}%
  \BibitemOpen
  \bibfield  {author} {\bibinfo {author} {\bibfnamefont {D~H}\ \bibnamefont
  {Slichter}}, \bibinfo {author} {\bibfnamefont {R}~\bibnamefont {Vijay}},
  \bibinfo {author} {\bibfnamefont {S.~J.}\ \bibnamefont {Weber}}, \bibinfo
  {author} {\bibfnamefont {S.}~\bibnamefont {Boutin}}, \bibinfo {author}
  {\bibfnamefont {Maxime}\ \bibnamefont {Boissonneault}}, \bibinfo {author}
  {\bibfnamefont {Jay~M}\ \bibnamefont {Gambetta}}, \bibinfo {author}
  {\bibfnamefont {Alexandre}\ \bibnamefont {Blais}}, \ and\ \bibinfo {author}
  {\bibfnamefont {Irfan}\ \bibnamefont {Siddiqi}},\ }\bibfield  {title}
  {\enquote {\bibinfo {title} {Measurement-induced qubit state mixing in
  circuit {QED} from upconverted dephasing noise},}\ }\href {\doibase
  10.1103/PhysRevLett.109.153601} {\bibfield  {journal} {\bibinfo  {journal}
  {Physical Review Letters}\ }\textbf {\bibinfo {volume} {109}},\ \bibinfo
  {pages} {153601} (\bibinfo {year} {2012})}\BibitemShut {NoStop}%
\bibitem [{\citenamefont {Sete}\ \emph {et~al.}(2014)\citenamefont {Sete},
  \citenamefont {Gambetta},\ and\ \citenamefont {Korotkov}}]{Sete:PRA:2014}%
  \BibitemOpen
  \bibfield  {author} {\bibinfo {author} {\bibfnamefont {Eyob~A}\ \bibnamefont
  {Sete}}, \bibinfo {author} {\bibfnamefont {Jay~M}\ \bibnamefont {Gambetta}},
  \ and\ \bibinfo {author} {\bibfnamefont {Alexander~N}\ \bibnamefont
  {Korotkov}},\ }\bibfield  {title} {\enquote {\bibinfo {title} {Purcell effect
  with microwave drive: Suppression of qubit relaxation rate},}\ }\href
  {\doibase 10.1103/PhysRevB.89.104516} {\bibfield  {journal} {\bibinfo
  {journal} {Physical Review B}\ }\textbf {\bibinfo {volume} {89}},\ \bibinfo
  {pages} {104516} (\bibinfo {year} {2014})}\BibitemShut {NoStop}%
\bibitem [{\citenamefont {Slichter}\ \emph {et~al.}(2016)\citenamefont
  {Slichter}, \citenamefont {M{\"u}ller}, \citenamefont {Vijay}, \citenamefont
  {Weber}, \citenamefont {Blais},\ and\ \citenamefont
  {Siddiqi}}]{Slichter:NJP:2016}%
  \BibitemOpen
  \bibfield  {author} {\bibinfo {author} {\bibfnamefont {D~H}\ \bibnamefont
  {Slichter}}, \bibinfo {author} {\bibfnamefont {Clemens}\ \bibnamefont
  {M{\"u}ller}}, \bibinfo {author} {\bibfnamefont {R}~\bibnamefont {Vijay}},
  \bibinfo {author} {\bibfnamefont {S.~J.}\ \bibnamefont {Weber}}, \bibinfo
  {author} {\bibfnamefont {Alexandre}\ \bibnamefont {Blais}}, \ and\ \bibinfo
  {author} {\bibfnamefont {Irfan}\ \bibnamefont {Siddiqi}},\ }\bibfield
  {title} {\enquote {\bibinfo {title} {Quantum zeno effect in the strong
  measurement regime of circuit quantum electrodynamics},}\ }\href {\doibase
  10.1088/1367-2630/18/5/053031} {\bibfield  {journal} {\bibinfo  {journal}
  {New Journal of Physics}\ }\textbf {\bibinfo {volume} {18}},\ \bibinfo
  {pages} {053031} (\bibinfo {year} {2016})}\BibitemShut {NoStop}%
\bibitem [{\citenamefont {Minev}\ \emph {et~al.}(2019)\citenamefont {Minev},
  \citenamefont {Mundhada}, \citenamefont {Shankar}, \citenamefont {Reinhold},
  \citenamefont {Guti{\'e}rrez-J{\'a}uregui}, \citenamefont {Schoelkopf},
  \citenamefont {Mirrahimi}, \citenamefont {Carmichael},\ and\ \citenamefont
  {Devoret}}]{Minev:N:2019}%
  \BibitemOpen
  \bibfield  {author} {\bibinfo {author} {\bibfnamefont {Z~K}\ \bibnamefont
  {Minev}}, \bibinfo {author} {\bibfnamefont {S~O}\ \bibnamefont {Mundhada}},
  \bibinfo {author} {\bibfnamefont {S}~\bibnamefont {Shankar}}, \bibinfo
  {author} {\bibfnamefont {P}~\bibnamefont {Reinhold}}, \bibinfo {author}
  {\bibfnamefont {R}~\bibnamefont {Guti{\'e}rrez-J{\'a}uregui}}, \bibinfo
  {author} {\bibfnamefont {Robert~J}\ \bibnamefont {Schoelkopf}}, \bibinfo
  {author} {\bibfnamefont {Mazyar}\ \bibnamefont {Mirrahimi}}, \bibinfo
  {author} {\bibfnamefont {H~J}\ \bibnamefont {Carmichael}}, \ and\ \bibinfo
  {author} {\bibfnamefont {Michel~H}\ \bibnamefont {Devoret}},\ }\bibfield
  {title} {\enquote {\bibinfo {title} {{To catch and reverse a quantum jump
  mid-flight}},}\ }\href {\doibase 10.1038/s41586-019-1287-z} {\bibfield
  {journal} {\bibinfo  {journal} {Nature}\ }\textbf {\bibinfo {volume} {570}},\
  \bibinfo {pages} {200--204} (\bibinfo {year} {2019})}\BibitemShut {NoStop}%
\bibitem [{\citenamefont {Petrescu}\ \emph {et~al.}(2020)\citenamefont
  {Petrescu}, \citenamefont {Malekakhlagh},\ and\ \citenamefont
  {T{\"u}reci}}]{Petrescu:PRB:2020}%
  \BibitemOpen
  \bibfield  {author} {\bibinfo {author} {\bibfnamefont {Alexandru}\
  \bibnamefont {Petrescu}}, \bibinfo {author} {\bibfnamefont {Moein}\
  \bibnamefont {Malekakhlagh}}, \ and\ \bibinfo {author} {\bibfnamefont
  {Hakan~E.}\ \bibnamefont {T{\"u}reci}},\ }\bibfield  {title} {\enquote
  {\bibinfo {title} {{Lifetime renormalization of driven weakly anharmonic
  superconducting qubits: {II}. The readout problem}},}\ }\href {\doibase
  10.1103/PhysRevB.101.134510} {\bibfield  {journal} {\bibinfo  {journal}
  {Physical Review B}\ }\textbf {\bibinfo {volume} {101}},\ \bibinfo {pages}
  {134510} (\bibinfo {year} {2020})}\BibitemShut {NoStop}%
\bibitem [{Note1()}]{Note1}%
  \BibitemOpen
  \bibinfo {note} {In the strong driving case, the displacement transformation
  should actually be applied first, before other perturbative expansions are
  performed. This would change some of the frequency dependence in the
  dissipators without changing their qualitative behaviour. For the purpose of
  simplicity we do not consider this additional complication here.}\BibitemShut
  {Stop}%
\end{thebibliography}%

\appendix

\section{Treatment of environmental terms\label{app:Bath}}

We consider three independent baths with their spectral functions defined through
\begin{align}
	\mean{\op o(t) \op o(t')} = \frac1\pi \int_{-\infty}^{\infty} d\omega\: \ee^{-\ii \omega (t-t')} C_{\op o}(\omega) \,,
\end{align}
with $\op o \in \{\op X, \op Z, \op R\}$. We do not consider any cross-terms in the expectation values, i.e. we set $\mean{\op X \op Z} = 0$ and similar.
At the same time we neglect any odd order bath expectation values, effectively assuming $\mean{\op o} = 0$.
The latter is synonymous with assuming that renormalisation of system energies due to the interaction with the bath is already taken into account in all Hamiltonian parameters. 
Further we make the Markov approximation, in that we assume that all baths remain in their thermal equilibrium at all times, 
and any disturbance of their equilibrium state decays with time-constants much faster than all system dynamics.

For a bosonic bath at thermal equilibrium, we find for the spectral function 
\begin{align}
	C_{\op o}(\omega) = \frac12 J(\abs\omega) \l( \coth{\frac{\abs\omega}{2T}} + \text{sign}(\omega) \r)
\end{align}
which follows detailed balance, $C_{\op o}(\omega) / C_{\op o}(-\omega) = \ee^{\omega/k_{B}T}$.
Here, $J(\omega)$ is the spectral density of the bath's coupling operator, defined in the continuum limit through
\begin{align}
	\sum_{k} \eta_{k}^{2} \rightarrow \frac1\pi \int_{0}^{\infty} d\omega\: \eta^{2}(\omega) \nu(\omega) = \frac1\pi \int_{0}^{\infty} d\omega\: J(\omega)
\end{align}
where we wrote the bath coupling operator as a sum over bosonic modes $\op o = \sum_{k} \eta_{k} \l( b_{k} + b_{k}\hc \r)$ and $\nu(\omega)$ is the bath's spectral density.

\section{Two level Rabi model\label{app:TwoLevel}}

Here we provide results for the special case of a two-level qubit, i.e. when $N=2$ in Eq.~\eqref{eq:H0}.
These can be obtained from the results in the main text by setting $g_{k} = \beta_{k} = \delta\omega_{k} = 0, \: \forall k>0$ 
and identifying the Pauli matrices as $\sigma_{z} = \sigma_{0,0} - \sigma_{1,1}$, $\sigma_{x} = \sigma_{1,0} + \sigma_{0,1}$.

The Hamiltonian is still written as $\H = \H_{0} + \H_{\text{int}} + \H_{\text{sys-env}} + \H_{\text{env},0}$ with
\begin{align}
	\H_{0} = \omega_{r} a\hc a - \frac12 \omega_{1,0} \sigma_{z} \,.
\end{align}
The interaction term in the Rabi model is now
\begin{align}
	\H_{\text{int, Rabi}} = g_{0} \sigma_{x} \l( a\hc + a \r) \,,
\end{align}
where we have transversal coupling between the qubit and resonator.
Making the rotating wave approximation in this coupling, we can instead write the Jaynes-Cummings interaction
\begin{align}
	\H_{\text{int, JC}} = g_{0} \left( \sigma_{-} a\hc + \sigma_{+} a \r)\,.
\end{align}
Finally, the system is coupled to three independent baths via
\begin{align}
	\H_{\text{sys-env}} = \sigma_{x} \op X + \sigma_{z} \op Z + \l( a\hc + a \r) \op R \,,
\end{align}
with the same conditions as previous on the bath operators $\op X, \op Z$, and $\op R$.
We again do not explicitly state the bath Hamiltonian $\H_{\text{env},0}$ here.

\subsection{Second order}

The dispersive shift for the Rabi interaction we write as
\begin{align}
	\H_{2, \text{Rabi}} = - \tilde\chi_{0} \l( \frac12\sigma_{z} + a\hc a \sigma_{z} \r) \,,
\end{align}
with $\tilde\chi_{0} = \frac{2 g_{0}^{2}\omega_{1,0}}{(\omega_{1,0}^{2}-\omega_{r}^{2})}$~\cite{Zueco:PRA:2009, Zhu:NJP:2013}
and for the Jaynes-Cummings interactions we find
\begin{align}
	\H_{2, \text{JC}} = -\chi_{0} \l( \frac12 \sigma_{z} + a\hc a \sigma_{z} \r) \,,
\end{align}
with the usual dispersive shift $\chi_{0} = \frac{g_{0}^{2}}{(\omega_{1,0}-\omega_{r})}$~\cite{Blais:PRA:2004}. 
Note that $\tilde\chi_{0} = 2\chi_{0} \frac{\omega_{1,0}}{\omega_{1,0}+\omega_{r}}$.
We also get the usual dissipative terms in the master equation at second order
\begin{align}
	\L_{2} \rho(t) =& \gamma_{\downarrow} \diss{\sigma_{-}}\rho(t) + \gamma_{\uparrow} \diss{\sigma_{+}}\rho(t) + \gamma_{\varphi}\diss{\sigma_{z}}\rho(t) \nn\\
		&+ \kappa_{-}\diss{a}\rho(t) + \kappa_{+}\diss{a\hc}\rho(t)
\end{align}
with the rates
\begin{align}
	\gamma_{\downarrow} &= C_{\op X}(\omega_{1,0})  \quad\,,\quad \gamma_{\uparrow} = C_{\op X}(-\omega_{1,0}) \,,\nn\\
	\gamma_{\varphi} &= C_{\op Z}(0) \,,\nn\\
	\kappa_{-} &= C_{\op R}(\omega_{r}) \quad\,,\quad \kappa_{+} = C_{\op R}(-\omega_{r}) \,.
\end{align}

\subsection{Fourth order}

At fourth order, we again focus on the correlated decay terms. We find the dominant contributions
\begin{align}
	\L_{4}\rho(t) =& \gamma_{\downarrow,P} \diss{\sigma_{-}}\rho(t) + \gamma_{\uparrow,P}\diss{\sigma_{+}}\rho(t) \nn\\
		&+ \gamma_{\downarrow+} \diss{\sigma_{-} a\hc}\rho(t) + \gamma_{\uparrow-}\diss{\sigma_{+}a}\rho(t) \nn\\
		&+ \gamma_{\downarrow -} \diss{\sigma_{-} a}\rho(t) + \gamma_{\uparrow+} \diss{\sigma_{+}a\hc}\rho(t) \nn\\
		&+\gamma_{\varphi-}\diss{\sigma_{z}a}\rho(t) + \gamma_{\varphi+}\diss{\sigma_{z} a\hc}\rho(t) \,,
\end{align}
with the Purcell rates
\begin{align}
	\gamma_{\downarrow,P} &= \frac{8 g_{0}^{2} \omega_{r}^{2}}{(\omega_{1,0}^{2} - \omega_{r}^{2})^{2} } C_{\op R}(\omega_{1,0}) \,,\nn\\ 
	\gamma_{\uparrow, P} &= \frac{8 g_{0}^{2} \omega_{r}^{2}}{(\omega_{1,0}^{2} - \omega_{r}^{2})^{2} } C_{\op R}(-\omega_{1,0}) \,,
\end{align}
the dressed dephasing contributions
\begin{align}
	\gamma_{\downarrow,+} &= \frac{8 g_{0}^{2}}{(\omega_{1,0} - \omega_{r})^{2}} C_{\op Z}(\omega_{1,0} - \omega_{r})\,,\nn\\
	\gamma_{\uparrow,-} &= \frac{8 g_{0}^{2}}{(\omega_{1,0} - \omega_{r})^{2}} C_{\op Z}(\omega_{r}- \omega_{1,0})\,,\nn\\
	\gamma_{\downarrow,-} &= \frac{8 g_{0}^{2}}{(\omega_{1,0} + \omega_{r})^{2}} C_{\op Z}(\omega_{1,0} + \omega_{r})\,,\nn\\
	\gamma_{\uparrow,+} &= \frac{8 g_{0}^{2}}{(\omega_{1,0} + \omega_{r})^{2}} C_{\op Z}(-\omega_{1,0} - \omega_{r})\,,
\end{align}
and the dephasing assisted gain and loss terms
\begin{align}
	\gamma_{\varphi,-} &= \frac{8 g_{0}^{2} \omega_{1,0}^{2}}{(\omega_{1,0}^{2} - \omega_{r}^{2})^{2} } C_{\op X}(\omega_{r})\,,\nn\\
	\gamma_{\varphi,+} &= \frac{8 g_{0}^{2} \omega_{1,0}^{2}}{(\omega_{1,0}^{2} - \omega_{r}^{2})^{2} } C_{\op X}(-\omega_{r})\,.
\end{align}

\section{Additional results \label{app:MoreResults}}

	\begin{figure}[htbp]
		\begin{center}
			\includegraphics[width=.95\columnwidth]{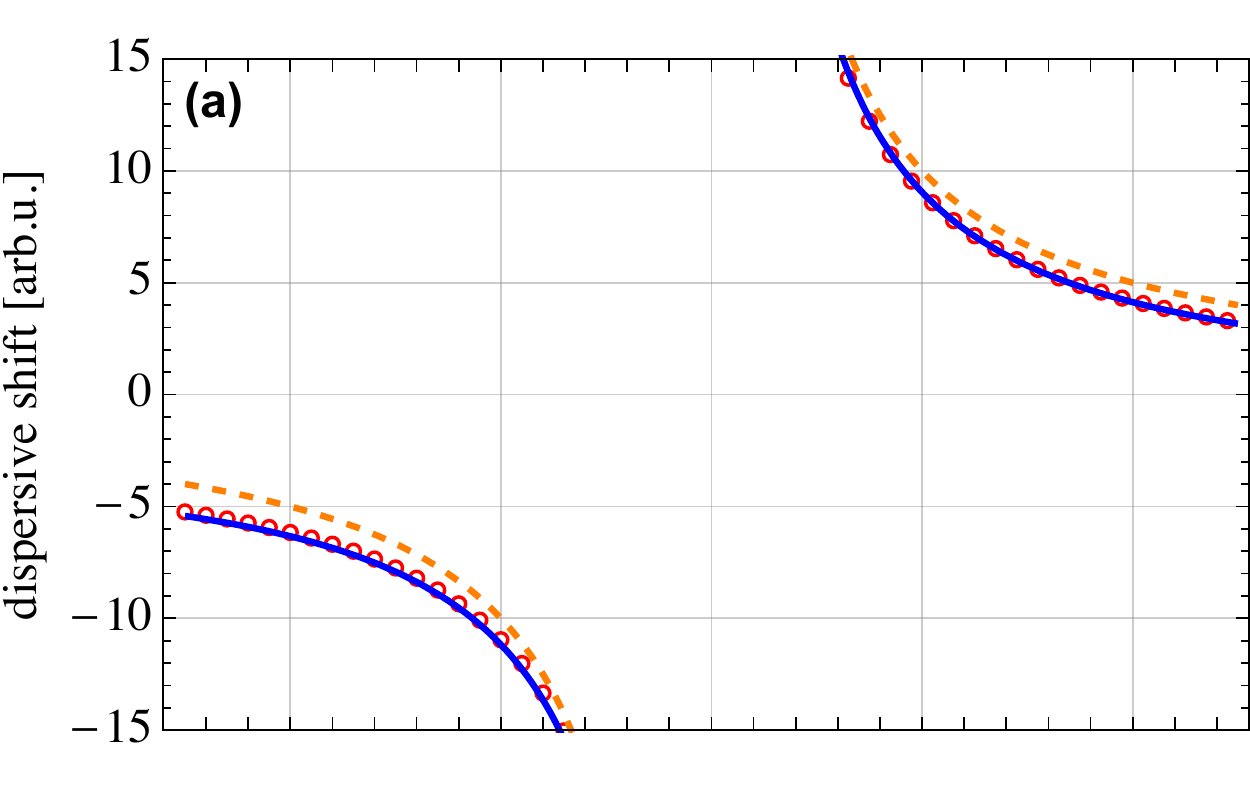}
			\includegraphics[width=.95\columnwidth]{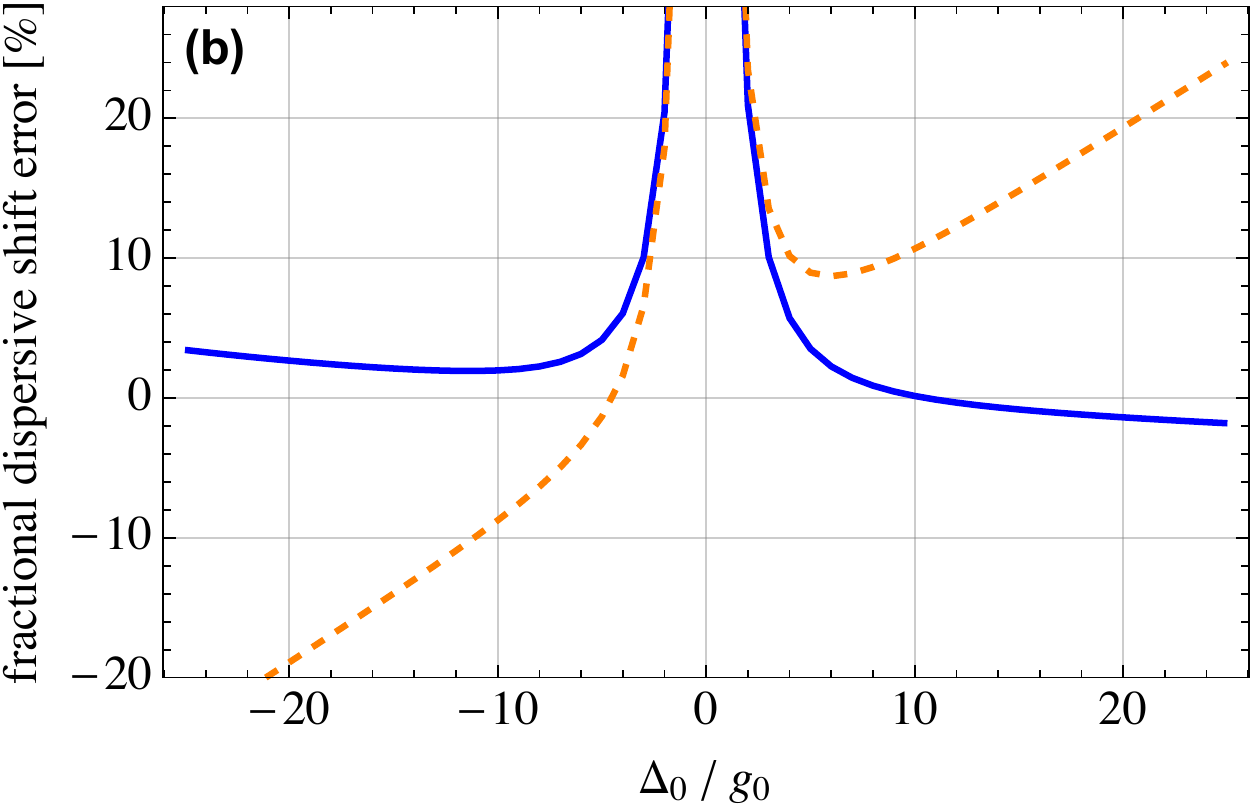}
			\caption{\textbf{(a)} Dispersive corrections to the first qubit level, given by $\chi_{0} - \xi_{1} + \xi_{0}$ when taking the Rabi interaction (solid blue) 
				and $\chi_{0}$ in the Jaynes-Cummings model (dashed orange).
				Red circles are again results from exact diagonalisation and \textbf{(b)} shows the error for both cases as a fraction of the exact diagonalisation result.
				Parameters as in Fig.~\ref{fig:DispersiveShift}, and additionally we have used the qubit anharmonicity $\alpha/2\pi = 250$~MHz.
				}
			\label{fig:DispersiveQ}
		\end{center}
	\end{figure}
	In the main text, we have shown the corrections to the one photon energy of the dispersive Hamiltonian, which is typically an experimentally readily available quantity. 
	Alternatively, experiments might track the qubit transitions frequency {between its ground and first excited state, while keeping the resonator in its ground state.} 
	In this case the corrections 
	are given by $\chi_{0} - \xi_{1} + \xi_{0}$ for the Rabi model and $\chi_{0}$ in the Jaynes-Cummings approximation.
	Fig~\ref{fig:DispersiveQ} shows a comparison of the Rabi and Jaynes-Cummings dispersive corrections for this case.
	Again, the expressions for the Rabi model are more accurate over nearly the full range of qubit-resonator detuning. 
	{In these calculations, as well as in the numerics underlying the exact diagonalisation results}, we assumed the multi-level qubit to be transmon-like, and have consequently replaced the higher level coupling strength with a harmonic approximation $g_{k} = \sqrt{k+1} \: g_{0}$ 
	and assumed a weak negative anharmonicity $\alpha$, with {$\omega_{k+1,k} = \omega_{1,0} - k\: \alpha$}.
	
\begin{figure}[htbp]
		\begin{center}
			\includegraphics[width=.95\columnwidth]{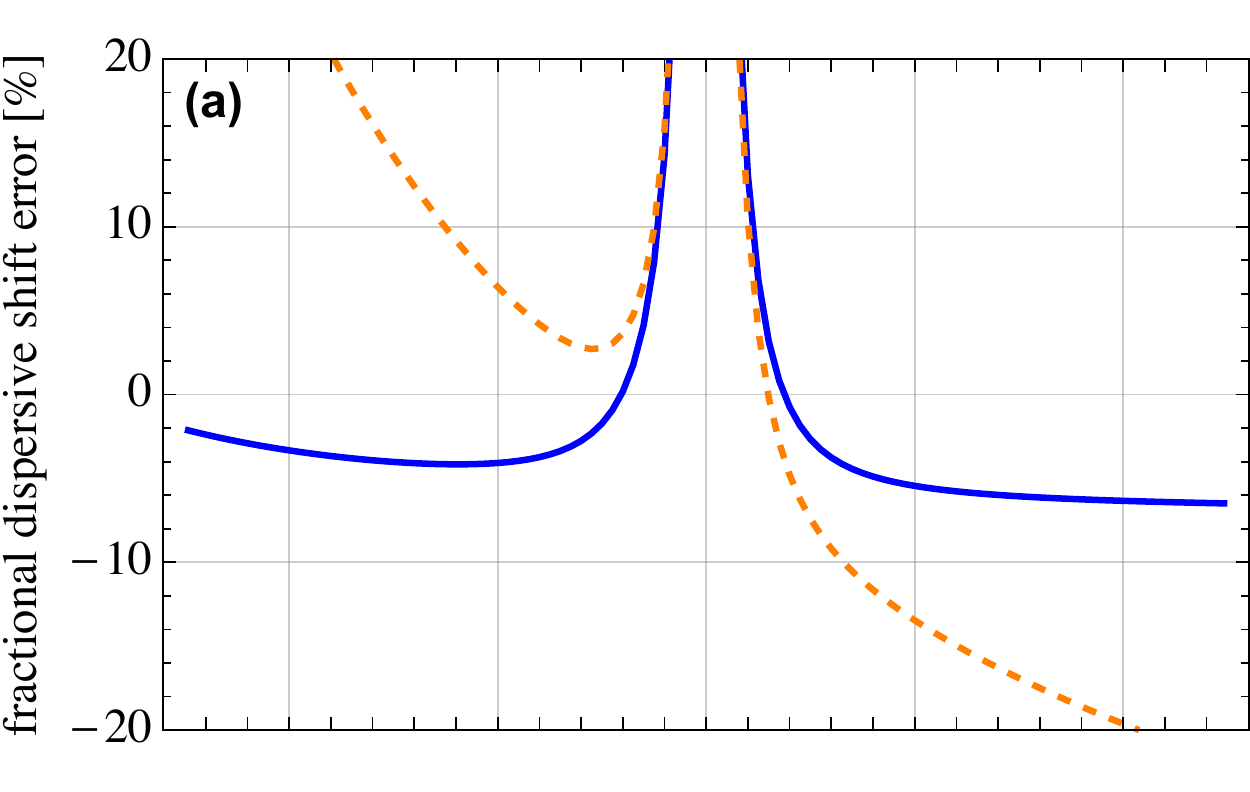}
			\includegraphics[width=.95\columnwidth]{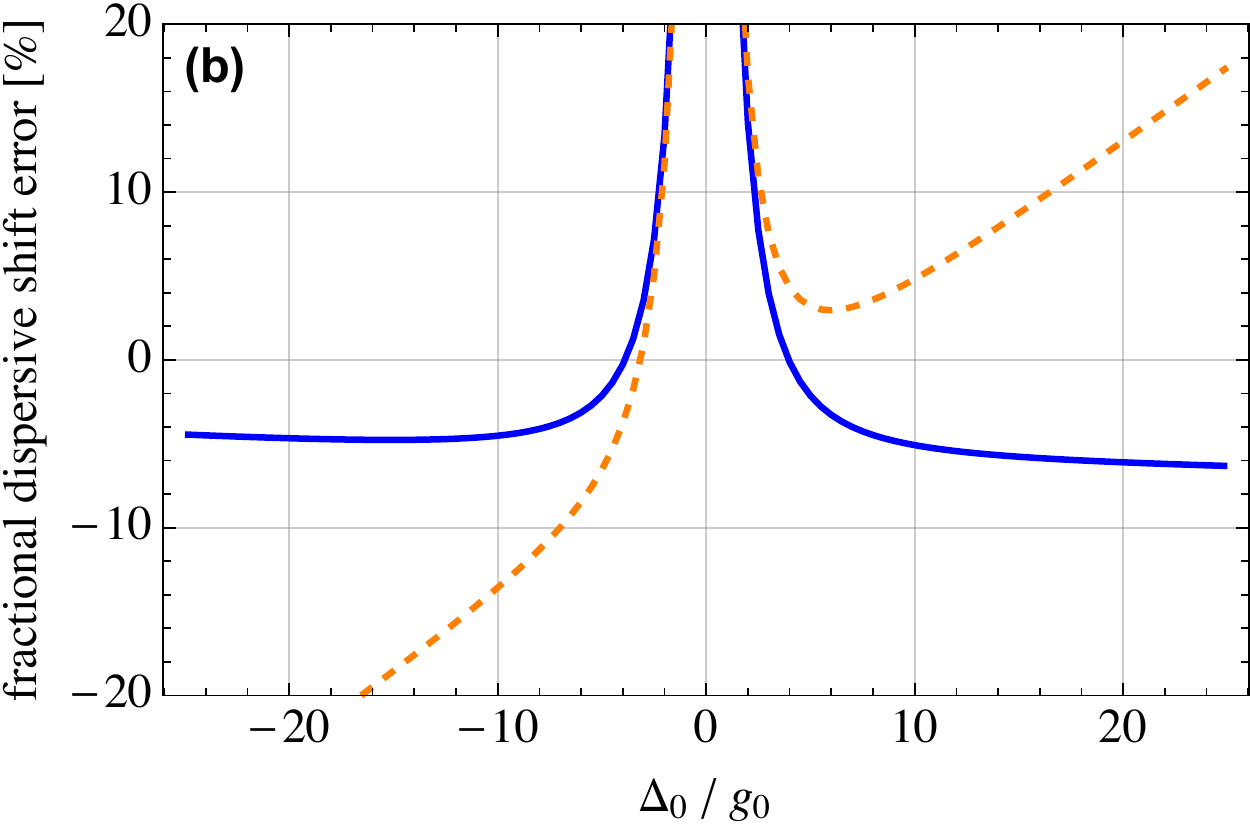}
			\caption{Fractional dispersive shift errors for \textbf{(a)} the resonator transition and \textbf{(b)} the qubit transition when treating the exact diagonalisation results as experimental data 
				and fitting the system coupling strength $g_{0}$ to this data.
				Parameters as in Fig.~\ref{fig:DispersiveQ}.
				}
			\label{fig:DispersiveErrorFit}
		\end{center}
	\end{figure}

	The results in Figs.~\ref{fig:DispersiveShift} and~\ref{fig:DispersiveQ} are calculated using the same numerical parameter values for both the exact numerics as well as the analytical approximations, 
	to demonstrate the quality of the perturbative approximations underlying the theory. 
	However, in experiments, the real Hamiltonian parameters are typically unknown, and need to be extracted using a model fit to measurement results. 
	To simulate this behavior, and to demonstrate that also for this case the analytical expression we derive here are performing well, we 
	treat the exact diagonalisation results from Fig.~\ref{fig:DispersiveShift} and~\ref{fig:DispersiveQ} 
	as input to a fit to the analytical dispersive shift expressions. Here we treat the system coupling strength $g_{0}$ as an unknown to be fitted.
	In Fig~\ref{fig:DispersiveErrorFit}, we then show the error as fraction of the exact numerical results for the analytical expressions for the Rabi (solid blue) and Jaynes-Cummings (dashed orange).
	Fit results for the system coupling strength $g$ are comparable for the two models. 
	For the fit to the resonator one photon energy, Fig~\ref{fig:DispersiveErrorFit}~\textbf{(a)}, we find $g_{0}/2\pi = 94.2 \pm 0.4$~MHz for a fit to the Rabi model results, 
	and $g_{0}/2\pi = 93.6 \pm0.9$~MHz for the expressions derived in Jaynes-Cummings model.
	Similarly, for the lowest qubit transition frequency fit, Fig.~\ref{fig:DispersiveErrorFit}~\textbf{(b)}, we find $g_{0}/2\pi = 94.3 \pm 0.4$~MHz in the Rabi model and 
	$g_{0}/ 2\pi = 94.7 \pm 0.9$~MHz in the Jaynes-Cummings case. 
	Therefore, any predictions one could make for the dissipative behavior of the system based on these results, 
	e.g. when calculating the ideal operating point in frequency taking into account Purcell and other incoherent corrections, 
	would then directly lead to the difference observed in the dissipative corrections in the main paper, i.e. Figs.~\ref{fig:Purcell} - \ref{fig:DephasingAssisted}.
 
\end{document}